\documentclass[12pt,aps,prd,preprint,tightenlines,
   showpacs,nofootinbib]{revtex4}
\newcommand{\PRE}[1]{{#1}} 

\usepackage{bm} \usepackage{epsfig}

\newcommand{\mweak}{M_{\text{weak}}}

\newcommand{\mgut}{M_{\text{GUT}}}
\newcommand{\mplanck}{M_{\text{Pl}}}
\newcommand{\mstar}{M_{*}}
\newcommand{\Omegachi}{\Omega_{\chi}}
\newcommand{\OmegaM}{\Omega_{\text{M}}}
\newcommand{\OmegaLambda}{\Omega_{\Lambda}}
\newcommand{\OmegaDM}{\Omega_{\text{DM}}}
\newcommand{\OmegaB}{\Omega_B}
\newcommand{\Mpc}{\text{Mpc}}
\newcommand{\ifb}{\text{fb}^{-1}}

\newcommand{\kev}{\text{keV}}

\newcommand{\gev}{\text{GeV}}
\newcommand{\tev}{\text{TeV}}

\newcommand{\m}{\text{m}}

\newcommand{\s}{\text{s}}
\newcommand{\yr}{\text{yr}}

\newcommand{\eqref}[1]{Eq.~(\ref{#1})}

\newcommand{\secref}[1]{Sec.~\ref{sec:#1}}

\newcommand{\figref}[1]{Fig.~\ref{fig:#1}}
\newcommand{\figsref}[2]{Figs.~\ref{fig:#1} and \ref{fig:#2}}

\newcommand{\NLSP}{\text{NLSP}}

\newcommand{\mNLSP}{m_{\NLSP}}
\newcommand{\mchi}{m_{\chi}}
\newcommand{\mgravitino}{m_{\gravitino}}

\newcommand{\gravitino}{\tilde{G}}

\newcommand{\mgaugino}{M_{1/2}}

\newcommand{\rem}[1]{{}}

\begin{document}

\preprint{UCI-TR-2005-35}

\title{
\PRE{\vspace*{1.5in}}
Dark Matter at the Fermi Scale
\PRE{\vspace*{0.3in}}
}

\author{Jonathan L.~Feng}
\affiliation{Department of Physics and Astronomy, University of
California, Irvine, CA 92697, USA
\PRE{\vspace*{.5in}}
}


\begin{abstract}
\PRE{\vspace*{.3in}} Recent breakthroughs in cosmology reveal that a
quarter of the Universe is composed of dark matter, but the
microscopic identity of dark matter remains a deep mystery.  I review
recent progress in resolving this puzzle, focusing on two
well-motivated classes of dark matter candidates: WIMPs and
superWIMPs.  These possibilities have similar motivations: they exist
in the same well-motivated particle physics models, the observed dark
matter relic density emerges naturally, and dark matter particles have
mass around 100 GeV, the energy scale identified as interesting over
70 years ago by Fermi.  At the same time, they have widely varying
implications for direct and indirect dark matter searches, particle
colliders, Big Bang nucleosynthesis, the cosmic microwave background,
and halo profiles and structure formation.  If WIMPs or superWIMPs are
a significant component of dark matter, we will soon be entering a
golden era in which dark matter will be studied through diverse probes
at the interface of particle physics, astroparticle physics, and
cosmology.  I outline a program of dark matter studies for each of
these scenarios, and discuss the prospects for identifying dark matter
in the coming years.
\end{abstract}


\maketitle

\section{Introduction \label{sec:intro}}

\begin{figure*}[t]
\centering
\includegraphics[height=4.0in]{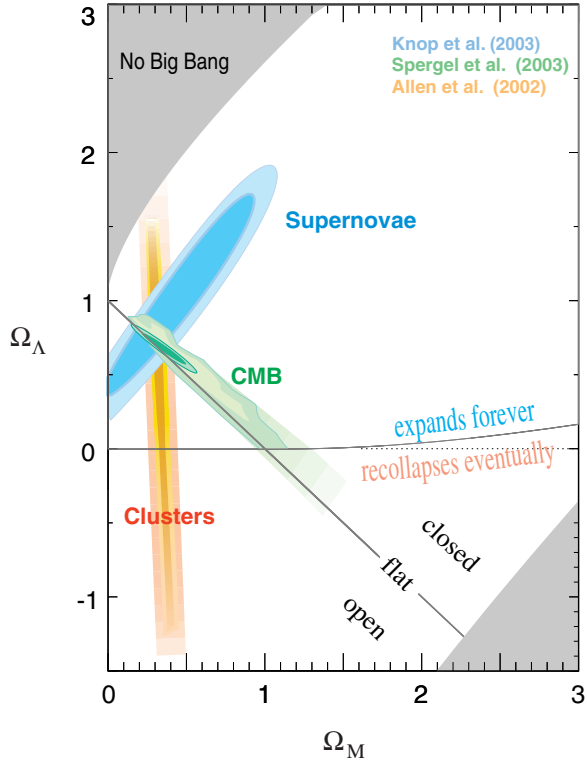}
\caption{Constraints on $\OmegaM$ and $\OmegaLambda$ from observations
  of supernovae, the CMB, and galaxy clusters~\cite{Knop:2003iy}.}
\label{fig:omegaplane}
\end{figure*}

In recent years, there has been tremendous progress in understanding
the Universe on the largest scales.  Observations of supernovae, the
cosmic microwave background (CMB), and galaxy clusters have provided
three stringent constraints on $\OmegaM$ and $\OmegaLambda$, the
energy densities of matter and dark energy in units of the critical
density.  These results are consistent and favor $(\OmegaM,
\OmegaLambda) \approx (0.3, 0.7)$, as shown in \figref{omegaplane}.
The amount of matter in the form of baryons is also constrained, both
by the CMB and by the observed abundances of light elements together
with the theory of Big Bang nucleosynthesis (BBN).  Although there are
at present possibly significant disagreements within the BBN data, the
CMB and BBN data taken as a whole are also impressively consistent,
providing yet another success for the current standard model of
cosmology.

Through these and many other observations, the total energy densities
of non-baryonic dark matter, baryons, and dark energy are constrained
to be~\cite{Spergel:2003cb,Tegmark:2003ud}
\begin{eqnarray}
\OmegaDM &=& 23\% \pm 4\% \\
\OmegaB &=& 4\% \pm 0.4\% \\
\OmegaLambda &=& 73\% \pm 4\% \ .
\end{eqnarray}
These results are remarkable.  At least two of the constraints of
\figref{omegaplane} must be wrong to change the conclusions about the
central values of $\OmegaM$ and $\OmegaLambda$.  These results are
also remarkably precise --- the fractional uncertainties on all three
are ${\cal O}(10\%)$.  Given that just a decade ago the range $0.2
\alt \OmegaDM \alt 0.6$ was allowed and $\OmegaLambda =0$ was often
assumed, this represents spectacular progress.  Although much of
cosmology remains imprecise, as we will see, the quantum leap in
precision in these three quantities already has dramatic implications
for particle physics.

At the same time, recent progress in cosmology is probably best viewed
as the first steps on the road to understanding the Universe.
Consider an historical precedent: in 200 B.C., Eratosthenes determined
the size of the Earth.  On a day when the Sun was directly overhead in
Syene, Eratosthenes sent a graduate student to measure the lengths of
shadows in Alexandria.  He was then able to extrapolate from the known
distance between these two cities to determine the circumference of
the Earth.  His answer was
\begin{equation}
2 \pi R_{\oplus} = 250,000~\text{stadia} \ .
\end{equation}
This result is remarkable.  At the time of publication, it was bigger
than many expected, leading many to be skeptical and helping to earn
Eratosthenes the nickname ``Beta''~\cite{Heath}.  His result was also
remarkably precise. We now know that it was good to less than
10\%~\cite{Goldstein,Rawlins1,Rawlins2}, where the leading source of
uncertainty is systematic error from the exact definition of the unit
``stadion''~\cite{Gulbekian}.  At the same time, the achievement of
Eratosthenes, though important, could hardly be characterized as a
complete understanding of the Earth.  Rather, it was just the
beginning of centuries of exploration, which eventually led to the
mapping of continents and oceans, giving us the picture of the Earth
we have today.

In a similar vein, recent breakthroughs in cosmology answer many
questions, but highlight even more.  Focusing on dark matter, the
primary subject of this review, these include
\begin{itemize}
\setlength{\baselineskip}{0.0in}
\item What particles form dark matter?
\item Is dark matter composed of one particle species or many?
\item What are dark matter's spin and other quantum numbers?
\item What are its interactions?
\item How and when was it produced?
\item Why does $\OmegaDM$ have the observed value?
\item How is dark matter distributed now?
\item What is its role in structure formation?
\item Is it absolutely stable?
\end{itemize}

\begin{figure*}[tb]
\centering
\includegraphics[width=0.95\textwidth]{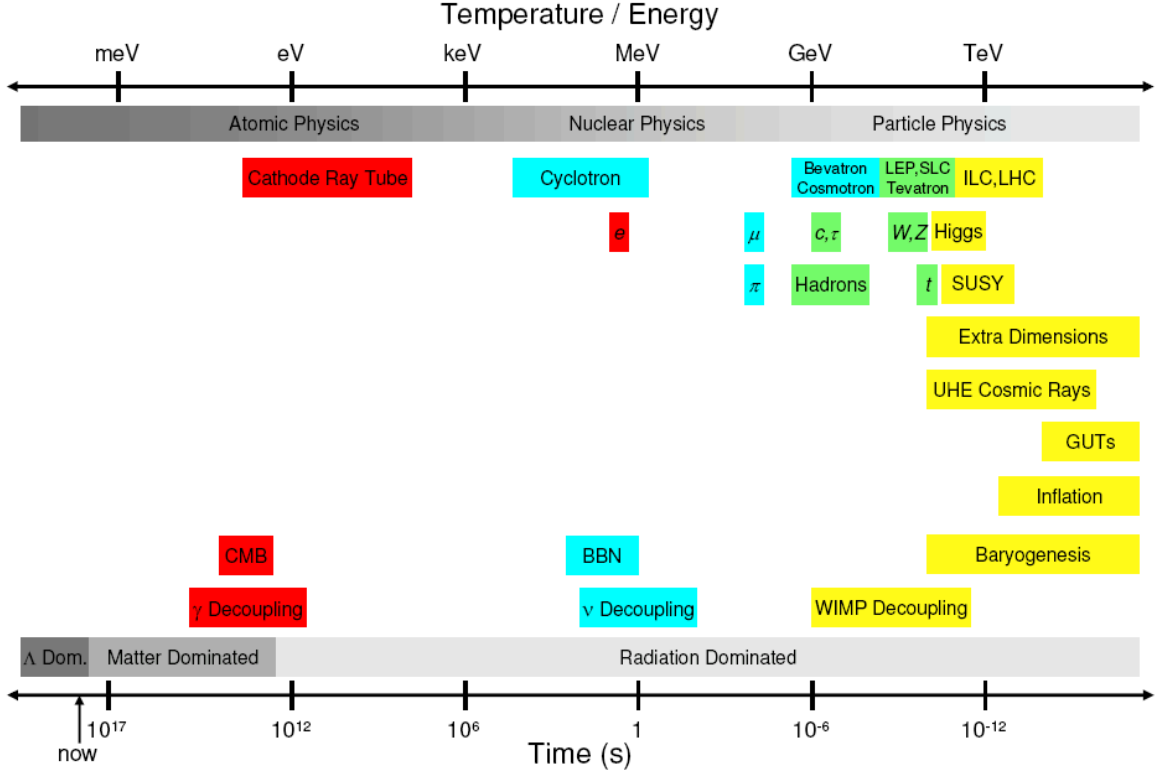}
\caption{Milestones in particle physics and cosmology along the
cosmological timeline.}
\label{fig:timeline}
\end{figure*}

Although these questions will continue to be sharpened by
astrophysical observations at large length scales, it is clear that
satisfying answers will require fundamental progress in our
understanding of microphysics.  This is nothing new --- the history of
advances in cosmology is to a large extent the story of successful
synergy between studies of the Universe on the smallest and largest
length scales.  This interplay is shown in \figref{timeline}, where
several milestones in particle physics and cosmology are placed along
the cosmological timeline.  As particle experiments reach smaller
length scales and higher energies, they probe times closer to the Big
Bang.  Just as atomic physics is required to interpret the CMB signal
from $t \sim 10^{13}~\s$ after the Big Bang and nuclear physics is
required to extrapolate back to BBN at $t \sim 1~\s$, particle
physics, and particularly the physics of the weak or Fermi scale, is
required to understand the era before $t \sim 10^{-8}~\s$, the era
that contains the answers to many of our most basic questions.

Here I will review recent progress in dark matter, focusing on new
proposals for and studies of dark matter at the Fermi scale
\begin{equation}
M_F \sim 100~\gev \ .
\end{equation}
$M_F$ is the scale of the weak interactions; its importance for
particle physics was realized by Fermi some 70 years ago.  Remarkably,
it is now appreciated that incisive studies of this scale may also
yield deep insights into the dark matter problem as well.  In the next
few years, the Tevatron and Large Hadron Collider (LHC) will play the
crucial role of opening the door to the Fermi scale. These will be
followed, we hope, by detailed studies of new physics at these
colliders and the proposed International Linear Collider (ILC).  If
dark matter particles do in fact have masses at the Fermi scale, these
colliders, in conjunction with cosmological and astroparticle probes,
will be essential for studying dark matter and unveiling its identity.

I begin with a review of the motivations to consider dark matter
candidates with Fermi scale masses.  These candidates may be divided
into two classes: WIMPs and superWIMPs.  I consider these in turn,
discussing the basic scenarios, their implications for particle
physics and cosmology, and the prospects for identifying dark matter
and answering many of the other questions listed above.  Further
background and discussion on both these and other dark matter topics
may be found in Refs.~\cite{reviews}.

\section{Why the Fermi Scale? \label{sec:fermi}}

The particle or particles that make up most of dark matter must be
stable, at least on cosmological time scales, and non-baryonic, so
that they do not disrupt the successes of BBN.  They must also be cold
or warm to properly seed structure formation, and their interactions
with normal matter must be weak enough to avoid violating current
bounds from dark matter searches.  The stringency of these criteria
pale in comparison with the unbridled enthusiasm of theorists, who
have proposed scores of viable candidates with masses and interaction
cross sections varying over tens of orders of magnitude.  In roughly
the order in which they were proposed, these include
axions~\cite{Peccei:1977ur,Wilczek:pj,Weinberg:1977ma},
thermally-produced gravitinos~\cite{Pagels:ke,Weinberg:zq,%
Krauss:1983ik,Nanopoulos:1983up,Khlopov:pf,Ellis:1984eq,Ellis:1984er,%
Juszkiewicz:gg}, neutralinos~\cite{Goldberg:1983nd,Ellis:1983ew},
axinos~\cite{Rajagopal:1990yx}, Q balls~\cite{Kusenko:1997si},
wimpzillas~\cite{Chung:1998ua}, self-interacting dark
matter~\cite{Spergel:1999mh}, annihilating dark
matter~\cite{Kaplinghat:2000vt}, Kaluza-Klein dark
matter~\cite{Servant:2002aq,Cheng:2002ej},
branons~\cite{Cembranos:2003mr,Cembranos:2003fu},
superWIMPs~\cite{Feng:2003xh,Feng:2003uy}, and many others.

Candidates with Fermi scale masses have received much of the attention,
however. There are at least four good reasons for this.  First, these
proposals are testable.  Second, new particles at the Fermi scale are
independently motivated by attempts to understand electroweak symmetry
breaking.  Third, these new particles often ``automatically'' have all
the right properties to be dark matter.  For example, their stability
often follows as a result of discrete symmetries that are necessary to
make electroweak theories viable, independent of cosmology.  And
fourth, these new particles are naturally produced with the
cosmological densities required of dark matter.

\begin{figure*}[t]
\centering
\includegraphics[height=2.7in]{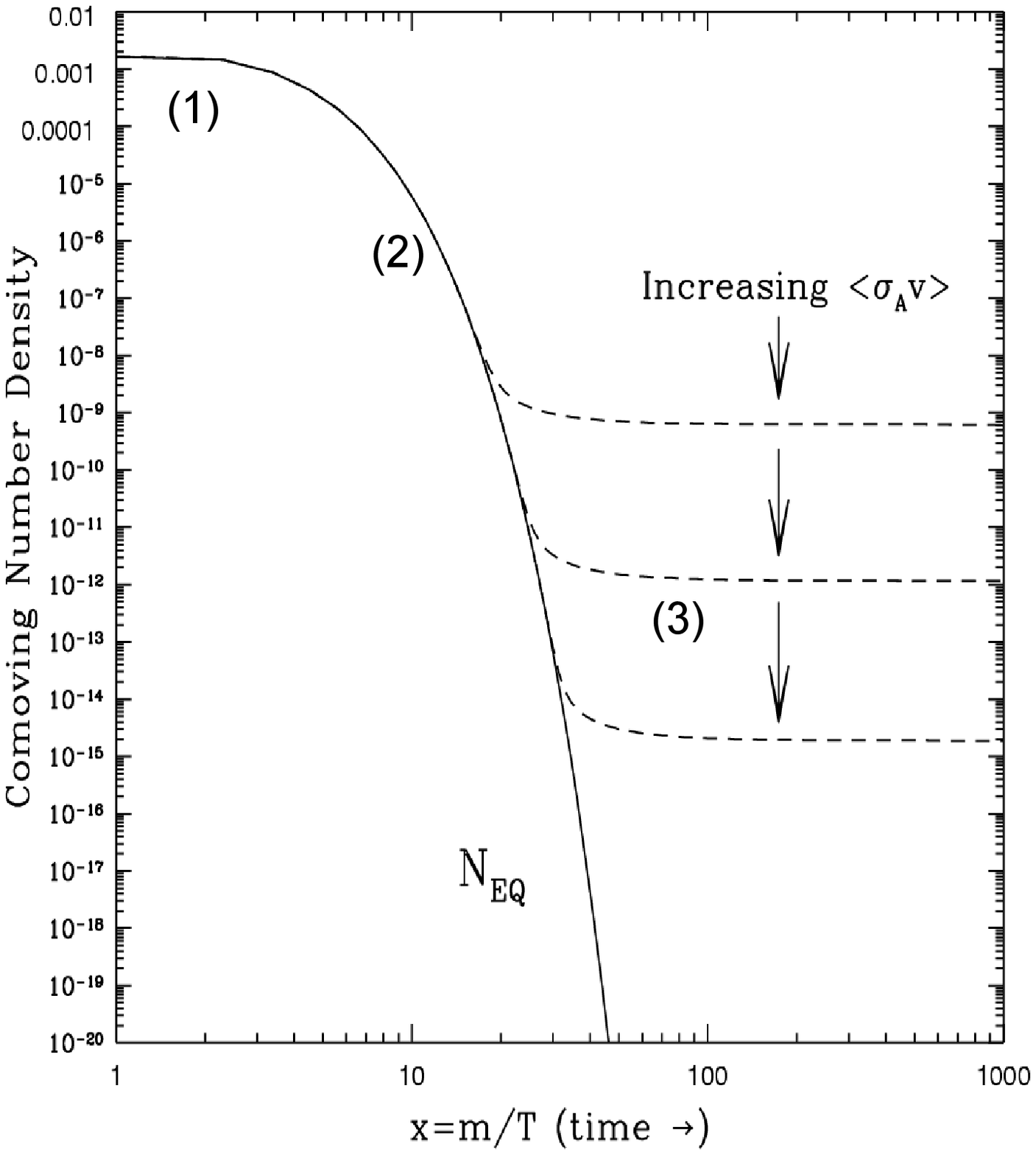}
\hspace*{0.2in}
\includegraphics[height=2.7in]{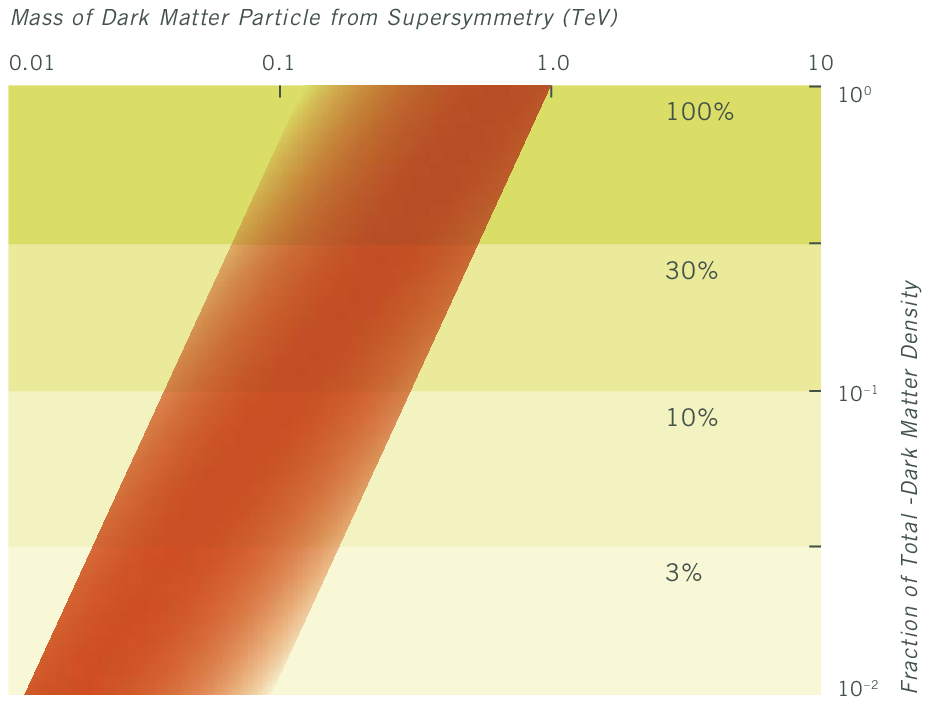}
\caption{Left: The cosmological evolution of a thermal relic's
  comoving number density. Right: A band of natural values in the
  $(\mchi, \Omegachi)$ plane for a thermal
  relic~\protect\cite{discovering}.}
\label{fig:freezeout}
\end{figure*}

The last motivation is particularly tantalizing.  Dark matter may be
produced in a simple and predictive manner as a thermal relic of the
Big Bang. The evolution of a thermal relic's number density is shown
in \figref{freezeout}. In stage (1), the early Universe is dense and
hot, and all particles are in thermal (chemical) equilibrium.  In
stage (2), the Universe cools to temperatures $T$ below the dark
matter particle's mass $\mchi$, and the number of dark matter
particles becomes Boltzmann suppressed, dropping exponentially as
$e^{-\mchi/T}$.  In stage (3), the Universe becomes so cold and dilute
that the dark matter annihilation rate is too low to maintain
equilibrium.  The dark matter particles then ``freeze out,'' with
their number asymptotically approaching a constant, their thermal
relic density.

More detailed analysis shows that the thermal relic density is rather
insensitive to $\mchi$ and inversely proportional to the annihilation
cross section: $\OmegaDM \sim \langle \sigma_A v \rangle^{-1}$.  The
constant of proportionality depends on the details of the
microphysics, but we may give a rough estimate.  On dimensional
grounds, the cross section can be written
\begin{equation}
\sigma_A v = k \frac{4 \pi \alpha_1^2}{m_{\chi}^2} \ 
(\text{1 or $v^2$})\ ,
\end{equation}
where $v$ is the relative velocity of the annihilating particles, the
factor $v^2$ is absent or present for $S$- or $P$-wave
annihilation, respectively, and terms higher-order in $v$ have been
neglected.  The constant $\alpha_1$ is the hypercharge fine structure
constant, and $k$ parameterizes deviations from this estimate.

With this parametrization, given a choice of $k$, the relic density is
determined as a function of $\mchi$.  The results are shown in
\figref{freezeout}.  The width of the band comes from considering both
$S$- and $P$-wave annihilation, and from letting $k$ vary from
$\frac{1}{2}$ to 2. We see that a particle that makes up all of dark
matter is predicted to have mass in the range $\mchi \sim 100~\gev -
1~\tev$; a particle that makes up 10\% of dark matter, still
significant (with respect to its impact on structure formation, for
example), has mass $\mchi \sim 30~\gev - 300~\gev$.  There are models
in which the effective $k$ is outside our illustrative range.  In
fact, values of $k$ smaller than we have assumed, predicting smaller
$\mchi$, are not uncommon, as the masses of virtual particles in
annihilation diagrams can be significantly higher than $\mchi$.  In
any case, the general conclusion remains: particles with mass at the
Fermi scale naturally have significant thermal relic densities.  For
this reason, a thorough exploration of the Fermi scale is crucial if
we hope to identify the particle or particles that make up dark
matter.  Even null results from LHC and ILC searches for dark matter
are important, as without them, it is unlikely that we will be able to
exclude the possibility of dark matter at the Fermi scale.

Given these results, many theories for new physics at the Fermi scale
contain promising dark matter candidates.  The candidates that exploit
the tantalizing numerical ``coincidence'' shown in \figref{freezeout}
may be grouped into two classes: WIMPs and superWIMPs. In the
following sections, we consider each of these two cases.

\section{WIMPs \label{sec:wimps}}

Weakly-interacting massive particles (WIMPs) have weak-scale masses
and weak-scale interactions.  They are an especially well-motivated
class of dark matter particles, and there are many examples, including
neutralinos in supersymmetry~\cite{Goldberg:1983nd,Ellis:1983ew},
Kaluza-Klein particles in theories with universal extra
dimensions~\cite{Servant:2002aq,Cheng:2002ej}, branons in theories
with large extra dimensions~\cite{Cembranos:2003mr,Cembranos:2003fu},
and the lightest $T$-odd particle in some little Higgs
theories~\cite{Cheng:2003ju}.

A program of detailed WIMP dark matter studies may be divided into
three (overlapping) stages:
\begin{enumerate}
\item WIMP Candidate Identification.  Is there evidence for WIMPs at
  colliders from, for example, events with missing energy and
  momentum?  Are there signals in dark matter search experiments?
  What are the candidates' masses, spins, and other quantum numbers?
\item WIMP Relic Density Determination. What are the dark matter
  candidates' predicted thermal relic densities?  Can they be
  significant components or all of dark matter?
\item Mapping the WIMP Universe. Combining collider results with
  results from direct and indirect dark matter searches, cosmological
  observations, and $N$-body simulations, what can we learn about
  astrophysical questions, such as structure formation and the
  distribution of dark matter in the Universe?
\end{enumerate}
Stage 1 is discussed in Ref.~\cite{whitepaper}.  In the following
sections, we will explore how well near future experiments may advance
Stages 2 and 3.

\begin{figure*}[t]
\centering
\includegraphics[height=2.8in]{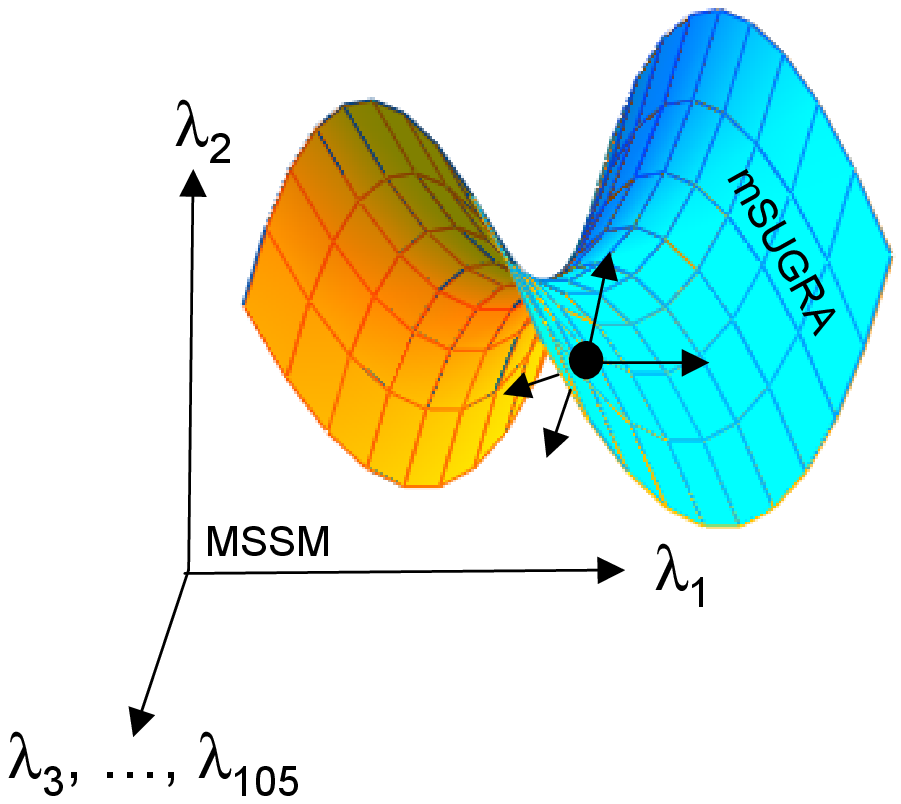}
\hspace*{0.2in}
\includegraphics[height=2.8in]{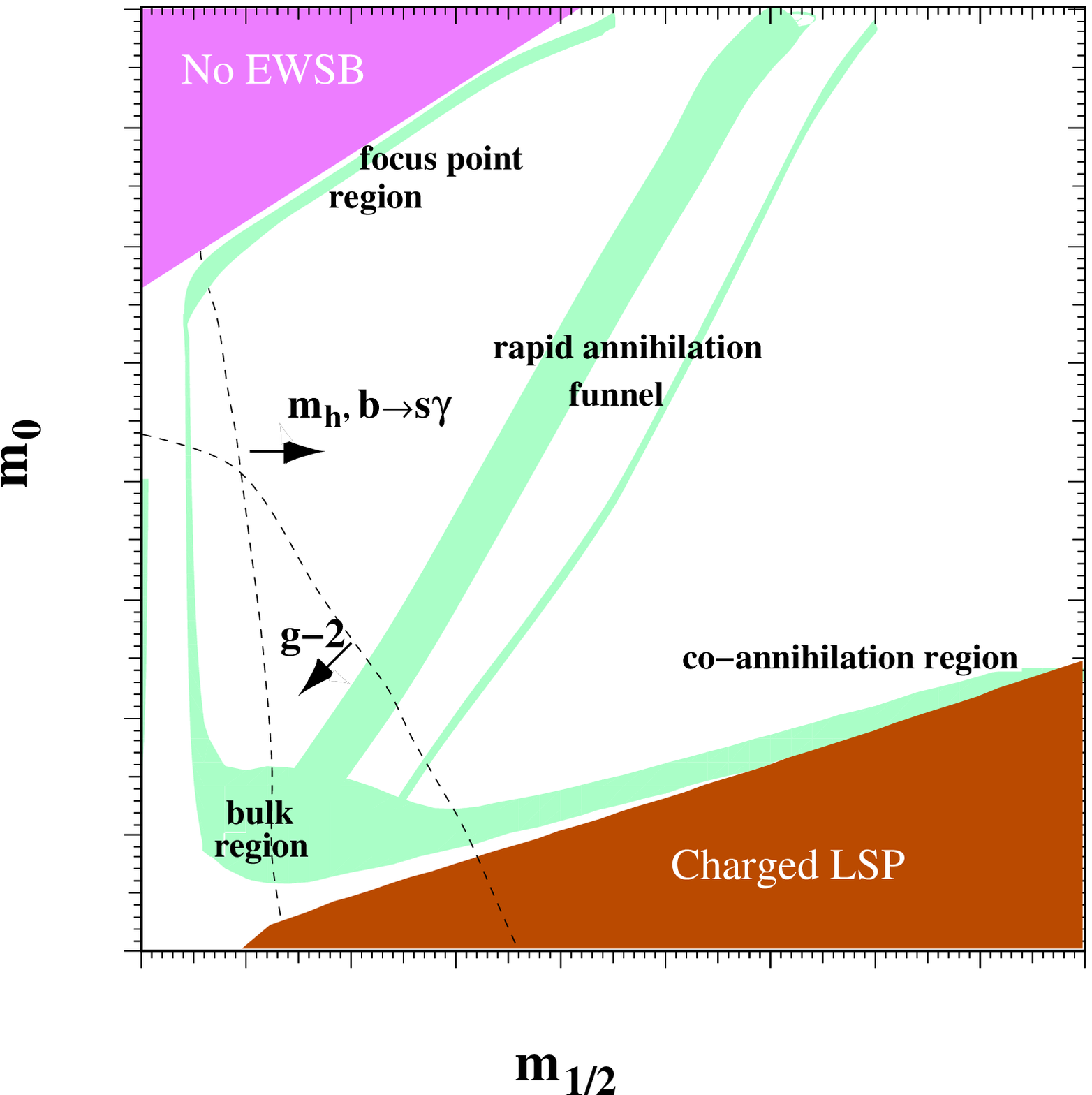}
\caption{Left: Minimal supergravity (mSUGRA) defines a hypersurface in
the 105-dimensional parameter space of the MSSM.  In the studies
described here, the underlying parameters are assumed to lie on the
mSUGRA hypersurface, but deviations in all directions of the MSSM
parameter space are allowed when evaluating the potential of colliders
to constrain parameters.  Right: Schematic diagram of regions with the
right amount of dark matter (shaded) in mSUGRA. This diagram is
qualitative.  The precise locations of the shaded regions depend on
suppressed parameters, and axis labels are purposely
omitted~\cite{whitepaper}. }
\label{fig:parameters}
\end{figure*}

To address these issues concretely, it is necessary to focus on one
representative example.  We choose to study neutralinos. Even with
this restriction, there are many qualitatively different scenarios. A
common choice is to study minimal supergravity (mSUGRA), a simple
model framework that encompasses many different possibilities.  In
this case, one assumes that the underlying supersymmetry parameters
realized in nature are those of a point in mSUGRA parameter space.  In
determining the capabilities of experiments, however, it is best to
relax all mSUGRA assumptions and ask how well the 105 parameters of
the general Minimal Supersymmetric Standard Model (MSSM) may be
determined.  This approach is illustrated in \figref{parameters}.

In much of mSUGRA parameter space the neutralino relic density lies
above the narrow allowed window, and so these possibilities are
cosmologically excluded.  The regions in which the neutralino relic
density is not too large, but is still sufficient to be all of dark
matter, are cosmologically preferred.  They have been given names and
include the bulk, focus point, co-annihilation, and rapid annihilation
funnel regions shown in \figref{parameters}.  Results from
representative models in each of the first two regions are summarized
below.  For results for the other two regions and related studies, see
Refs.~\cite{Gray:2005ci,Birkedal:2005jq,Battaglia:2005ie,%
Moroi:2005nc,Allanach:2004xn}.  For each model, the superpartner
spectrum is determined by the computer code
ISAJET~\cite{Paige:2003mg}, and cosmological observables, such as the
thermal relic density, are determined by the software packages
DARKSUSY~\cite{Gondolo:2004sc} and micrOMEGAs~\cite{Belanger:2004yn}.

\subsection{WIMP Relic Density Determination \label{sec:relic}}

\begin{figure*}[t]
\centering
\includegraphics[width=0.95\textwidth]{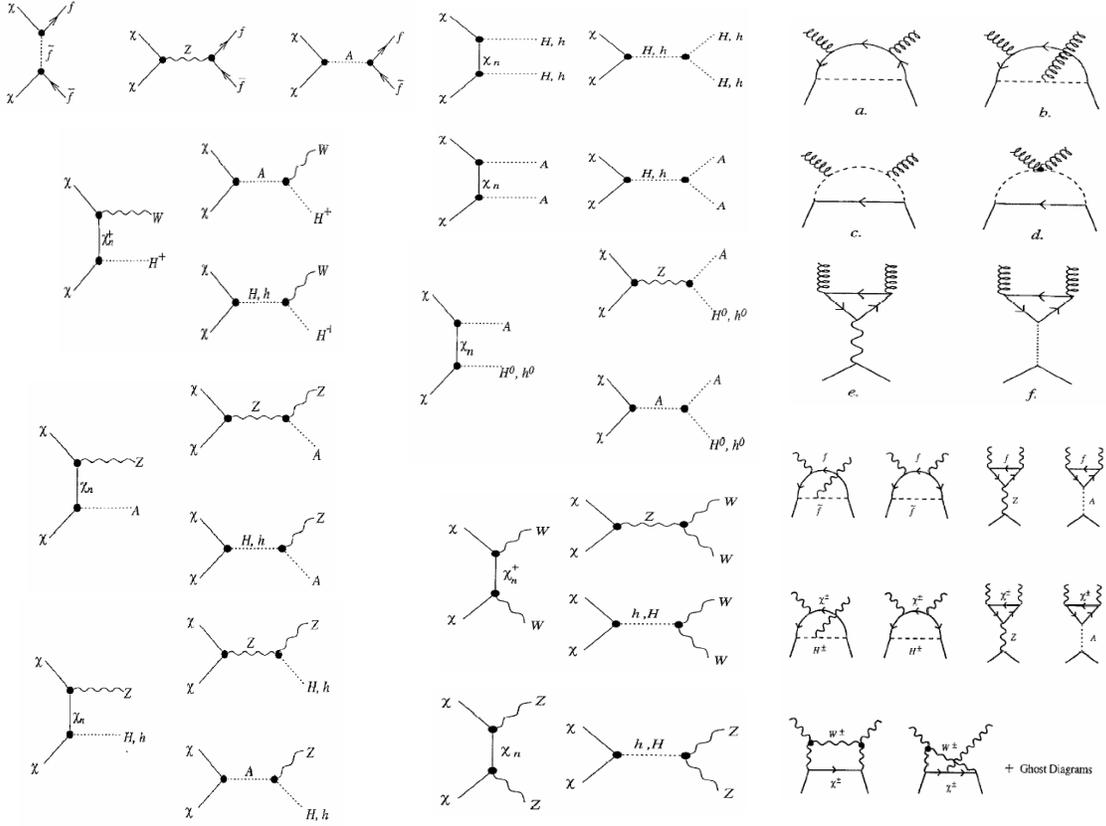}
\caption{Processes contributing to neutralino WIMP
annihilation~\cite{anngraphs}.
}
\label{fig:annihilationgraphs}
\end{figure*}

To determine the predicted WIMP thermal relic density, one must
experimentally constrain all processes contributing significantly to
the WIMP pair annihilation cross section.  This requires detailed
knowledge not only of WIMPs and their properties, but also of all
other particles contributing to their annihilation.  This is no small
task --- the number of processes contributing to WIMP annihilation is
large, as illustrated in \figref{annihilationgraphs}.  All unknown
parameters at the Fermi scale must be either measured precisely or
constrained sufficiently so that their effects are known to be
irrelevant.  Such detailed work relies primarily on particle
colliders, and we now consider how well colliders may constrain the
thermal relic density.

\subsubsection{Bulk Region}

In the bulk region, a much studied model is specified by the mSUGRA
parameters of Linear Collider Cosmology Model 1 (LCC1):
\begin{equation}
\text{LCC1:} \ (m_0, M_{1/2}, A_0, \tan\beta) = 
(100~\gev, 250~\gev, -100~\gev, 10) \ ,
\end{equation}
with $\mu > 0$, $m_{3/2} > m_{\text{LSP}}$, and $m_t = 178~\gev$.
Here $m_0$, $M_{1/2}$, and $A_0$ are the universal scalar, gaugino,
and trilinear coupling masses specified at the grand unified scale
$\mgut \simeq 2.4 \times 10^{16}~\gev$, respectively, $\tan\beta
\equiv \langle H^0_u \rangle / \langle H^0_d \rangle$ is the ratio of
Higgs boson vacuum expectation values, $\mu$ is the supersymmetric
Higgs mass, and $m_{3/2}$ is the gravitino mass.  The neutralino
thermal relic density at this point is $\Omegachi h^2 = 0.19$ ($h
\simeq 0.71$), significantly higher than the range $\Omegachi h^2 =
0.113 \pm 0.009$ allowed by the latest cosmological
constraints~\cite{Spergel:2003cb,Tegmark:2003ud}.  Nevertheless, the
choice of LCC1 is convenient, since it has been studied in great
detail in other studies, where it is also known as Snowmass Points and
Slopes Model 1a (SPS1a)~\cite{Allanach:2002nj}.

\begin{figure*}[t]
\centering
\includegraphics[width=2.1in]{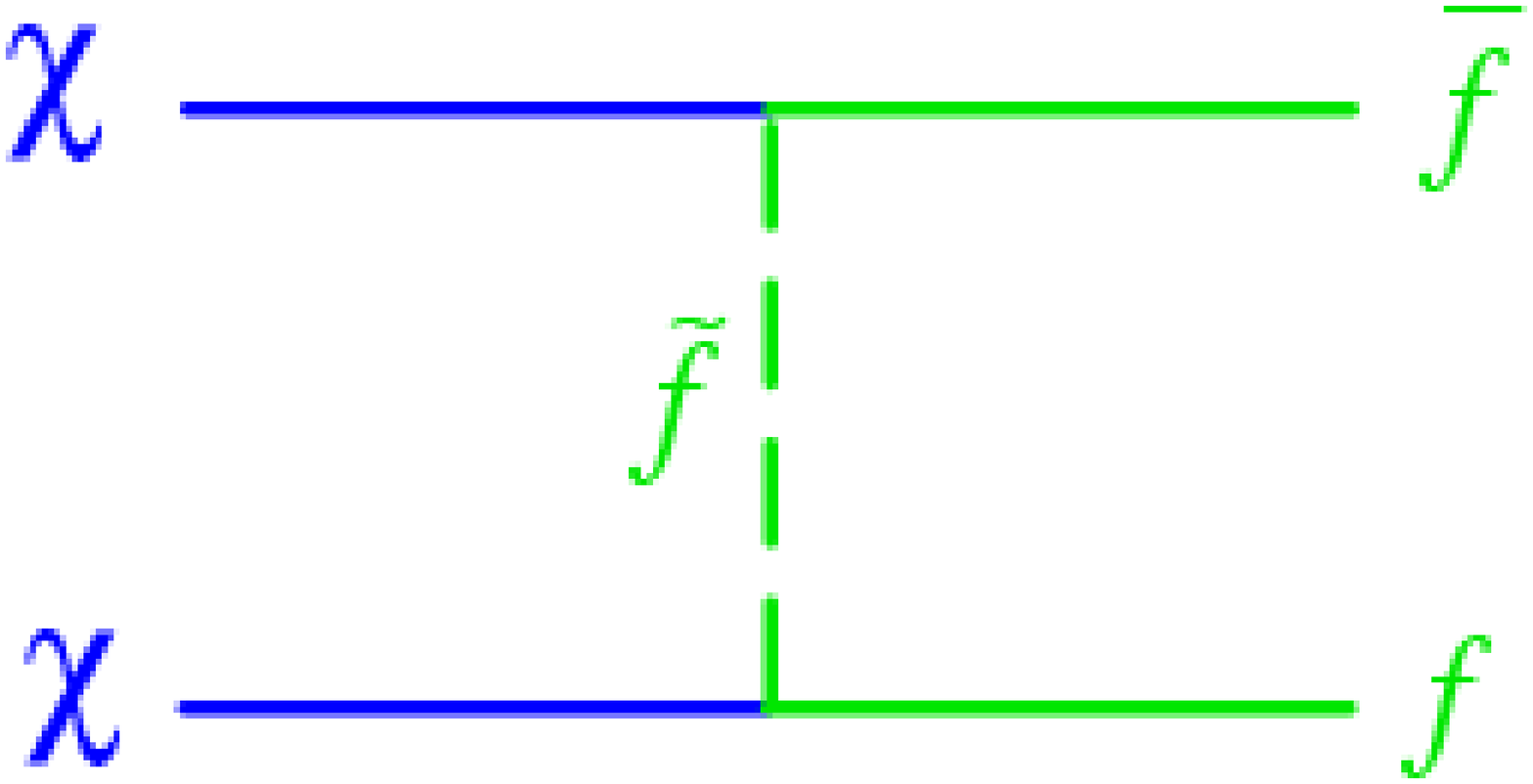}
\hspace*{0.2in}
\includegraphics[width=4.0in]{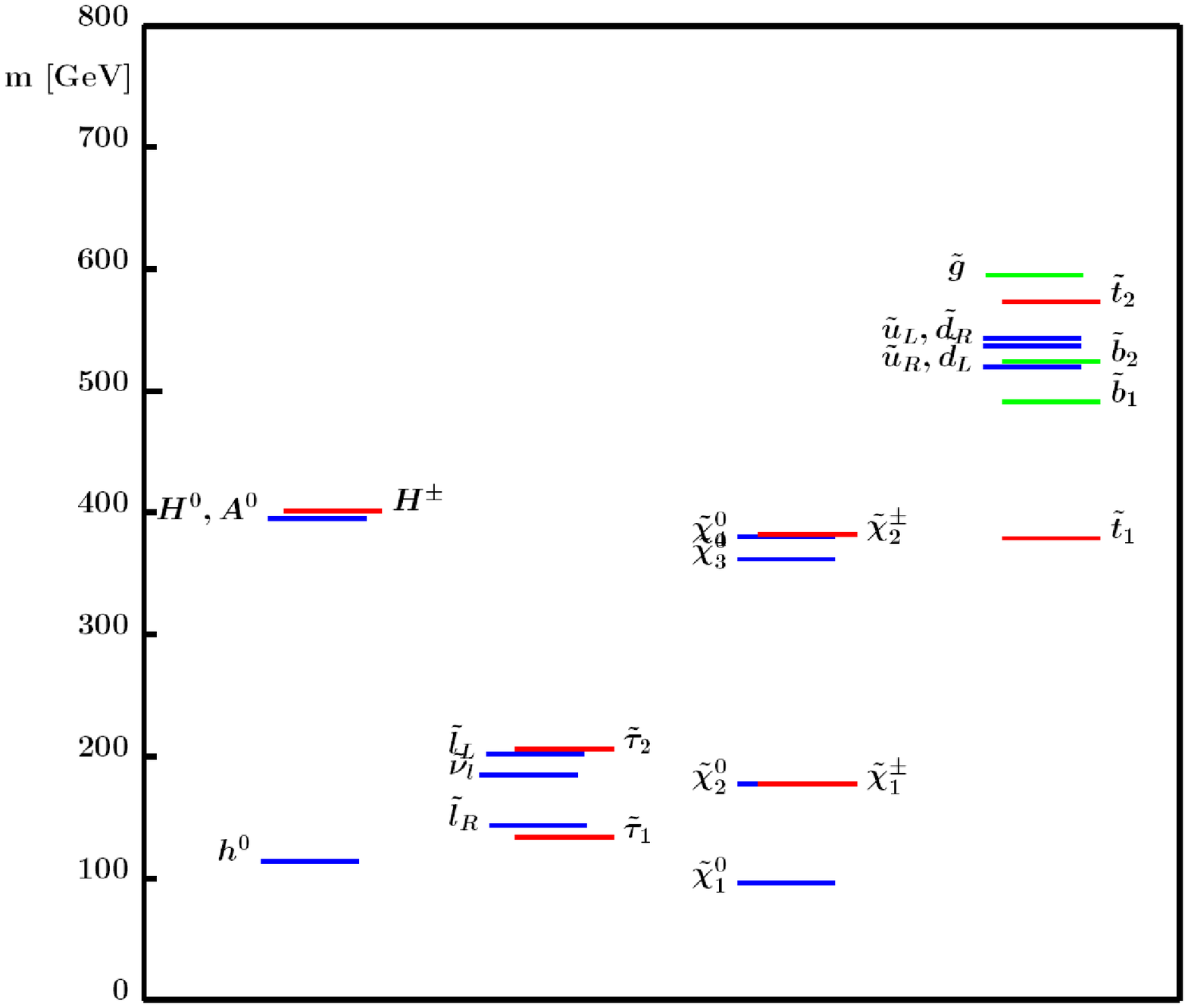}
\caption{Left: The dominant neutralino annihilation process in the
  bulk region.  Right: The superpartner spectrum at LCC1, a
  representative model in the bulk region~\cite{Allanach:2002nj}.}
\label{fig:bulk_feyn}
\end{figure*}

In the bulk region, neutralinos annihilate dominantly through $\chi
\chi \to f \bar{f}$ through a $t$-channel scalar $\tilde{f}$, as shown
in \figref{bulk_feyn}.  To achieve the correct relic density, this
process must be efficient, requiring light sfermions and, since the
neutralino must be the lightest supersymmetric particle (LSP), light
neutralinos.  These characteristics are exhibited in the superpartner
spectrum of LCC1, shown in \figref{bulk_feyn}.  It is noteworthy that
in this case, cosmology provides a strong motivation for light
superpartners within the reach of a 500 GeV ILC.

To determine the relic density at LCC1, all of the supersymmetry
parameters entering annihilation processes, including those shown in
\figref{bulk_feyn} and others, must be determined to high accuracy.
The LCC1 superpartner spectrum makes possible many high precision
measurements at the LHC.  LCC1 (SPS1a) is in significant respects a
``best case scenario'' for the LHC.  The implications of these
measurements for cosmology will be summarized below.

The LHC results may be improved at the ILC.  For example, superpartner
masses may be determined with extraordinary precision through
kinematic endpoints and threshold scans, as shown in
\figref{bulk_threshold}.  The kinematic endpoints of final state
leptons in the process $e^+ e^- \to \tilde{l}^+ \tilde{l}^- \to l^+
l^- \chi \chi$ determine both $\tilde{l}$ and $\chi$ masses.  Slepton
masses may also be determined through threshold scans.  Threshold
scans provide even higher precision, and may actually {\em save}
luminosity.  This is the case, for example, for selectron mass
determinations through $e^-e^-$ threshold scans, where precisions of
tens of MeV may be obtained with 1 to $10~\ifb$ of integrated
luminosity~\cite{Feng:1998ud,Feng:2001ce,Freitas:2003yp}.  More
generally, the required measurements exploit the full arsenal of the
ILC, from its variable beam energy, to its polarized beams, to the
$e^-e^-$ option.  The results of one study are summarized in
\figref{bulk_table}.

\begin{figure*}[t]
\centering
\includegraphics[height=2.1in]{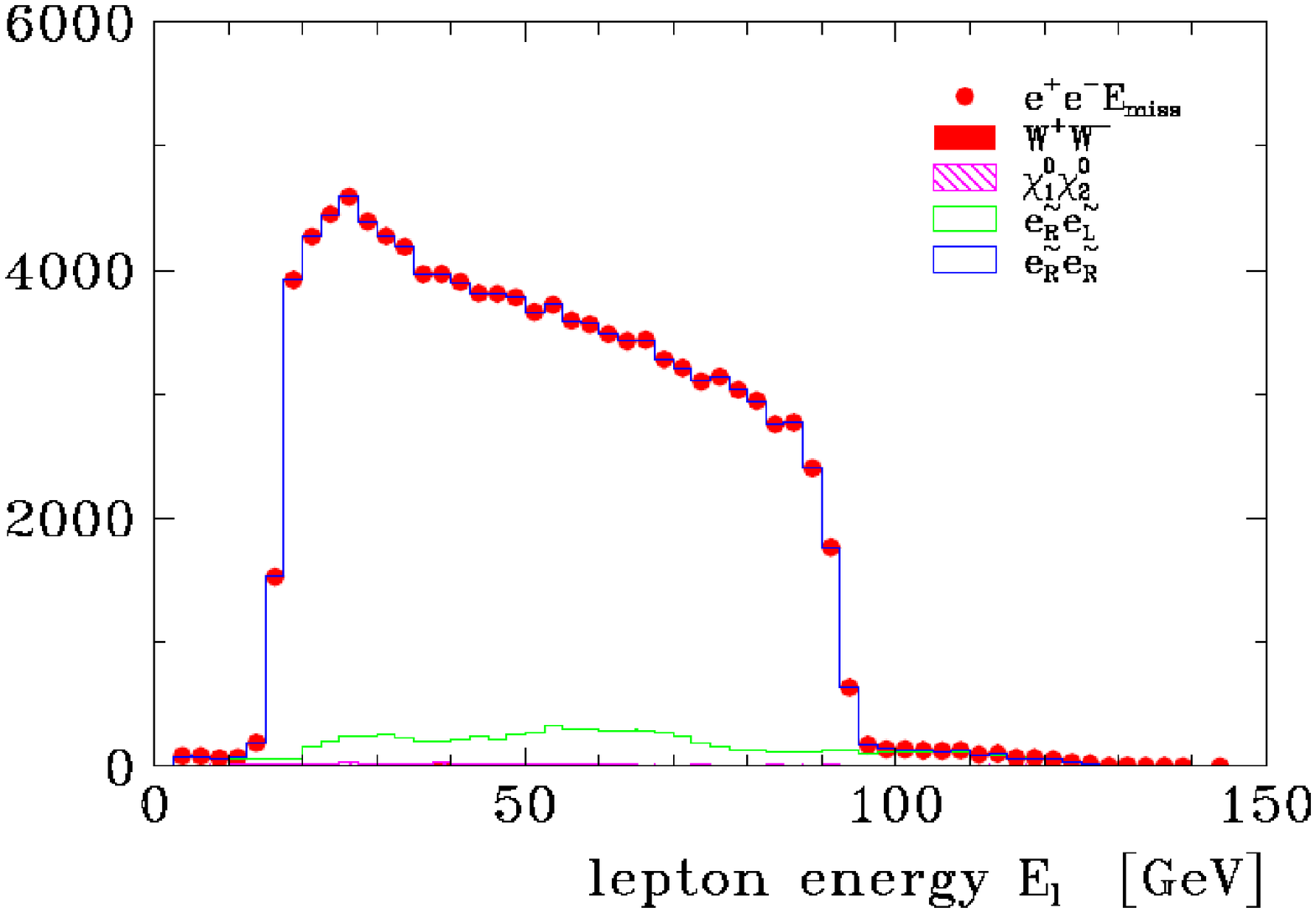}
\hspace*{0.2in}
\includegraphics[height=2.1in]{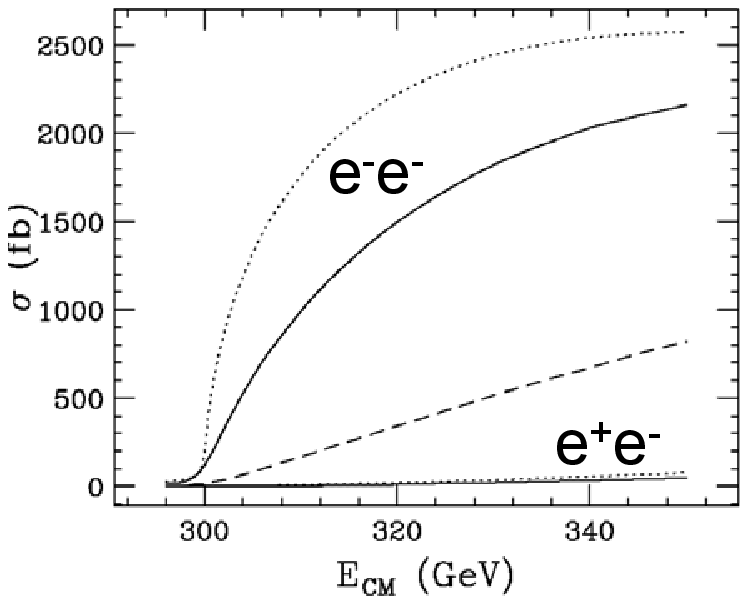}
\caption{Left: The kinematic endpoints of lepton energies from $e^+
e^- \to \tilde{l}^+ \tilde{l}^- \to l^+ l^- \chi \chi$ provide precise
determinations of slepton and neutralino
masses~\cite{Weiglein:2004hn}.  Right: Threshold scans may also be
used to determine slepton masses.  In the case of selectron masses,
$e^-e^-$ threshold scans provide higher precision and simultaneously
{\em save} luminosity~\cite{Feng:2001ce}.}
\label{fig:bulk_threshold}
\end{figure*}

\begin{figure*}[t]
\centering
\includegraphics[width=0.95\textwidth]{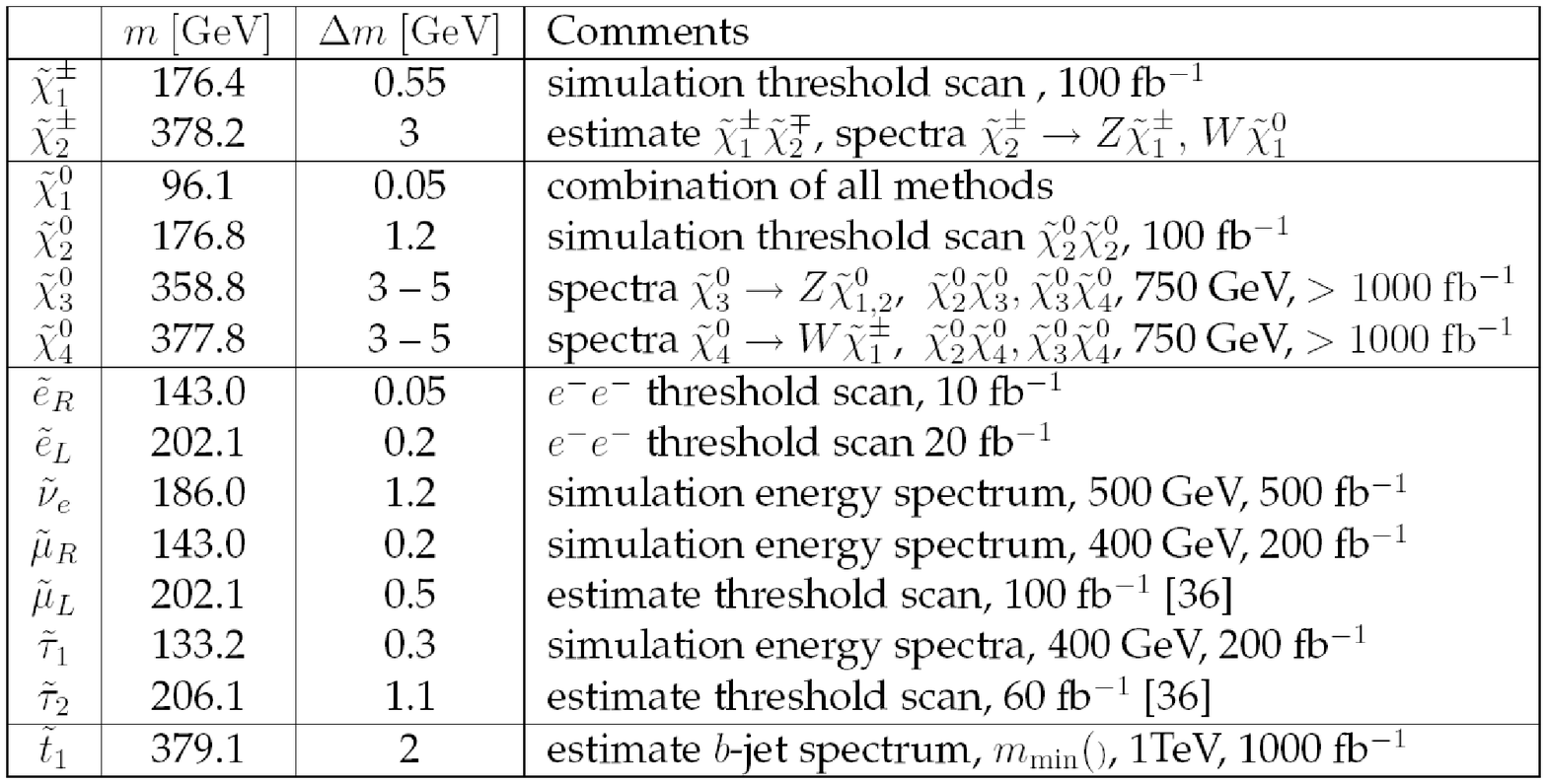}
\caption{The precision with which superpartner masses may be
  determined at the ILC for charginos $\tilde{\chi}^\pm$, neutralinos
  $\tilde{\chi}^0$, sleptons $\tilde{l}$, and top squark
  $\tilde{t}_1$.  The first column gives the underlying value of the
  masses, the second the constraint from collider studies, and the
  third the method used to achieve the
  constraint~\cite{Weiglein:2004hn}.}
\label{fig:bulk_table}
\end{figure*}

The neutralino thermal relic density may be determined by combining
the precise determination of all relevant supersymmetry parameters and
also verifying the insensitivity of the relic density to all other
parameters.  The results depend somewhat on the prescription one uses
to combine these data.  One approach is to choose points in parameter
space at random, weighting each with a Gaussian distribution for each
observable.  The relic density allowed region is then identified as
the symmetric interval around the central value that contains 68\% of
the weighted probability.

The result of applying this method with 50,000 model parameter points
randomly selected around LCC1 is shown in \figref{bulk_omega}.  The
result is that the ILC may determine the thermal relic density to a
fractional uncertainty of
\begin{equation}
\text{LCC1 (preliminary):} \ 
\frac{\Delta (\Omegachi h^2)}{\Omegachi h^2} = 2.2\% 
\quad [ \Delta (\Omegachi h^2) = 0.0042 ] \ .
\end{equation}
The current constraint from WMAP and the projected future constraint
from the Planck satellite are also shown.  WMAP and Planck provide no
information about the mass of the dark matter particle.

\begin{figure*}[t]
\centering
\includegraphics[width=0.8\textwidth]{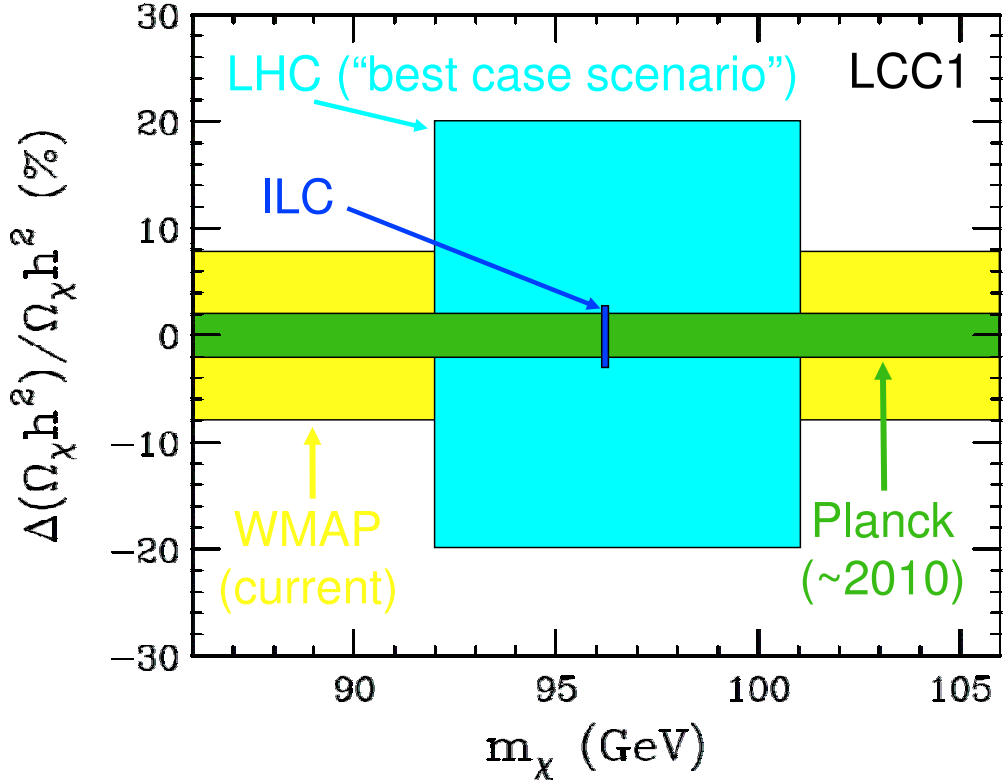}
\caption{Constraints in the $(\mchi, \Delta (\Omegachi h^2)/ \Omegachi
h^2 )$ plane from the ILC and LHC.  Constraints on $\Delta (\Omegachi
h^2)/ \Omegachi h^2$ from the WMAP and Planck satellite experiments
are also shown. The satellite experiments provide no constraints on
$\mchi$.}
\label{fig:bulk_omega}
\end{figure*}

\subsubsection{Focus Point Region}

In the focus point region, one may choose the representative model,
Linear Collider Cosmology Model 2, specified by
\begin{equation}
\text{LCC2:} \ (m_0, M_{1/2}, A_0, \tan\beta) = 
(3280~\gev, 300~\gev, 0, 10) \ ,
\end{equation}
with $\mu > 0$, $m_{3/2} > m_{\text{LSP}}$, $m_t = 175~\gev$.  In
focus point supersymmetry~\cite{Feng:1999hg,Feng:1999mn}, squarks and
sleptons are very heavy, and so the diagrams that are dominant in the
bulk region are suppressed.\footnote{As a result of this property,
models like focus point supersymmetry may be challenging for
supersymmetry discovery and study at the LHC~\cite{Baer:2005ky}.}
Nevertheless, the desired relic density may be
achieved~\cite{Feng:2000gh}, because in the focus point region, the
neutralino is not a pure Bino, but contains a significant Higgsino
component. The processes $\chi \chi \to W^+ W^-$, shown in
\figref{fp_feyn}, and $\chi \chi \to ZZ$, which are negligible in the
bulk region, therefore become efficient.  Neutralino mixing is
typically achieved when neutralinos and charginos are fairly light and
not too split in mass, and so the demands of neutralino dark matter
motivate supersymmetry with light neutralinos and charginos.

\begin{figure*}[t]
\centering \includegraphics[height=0.9in]{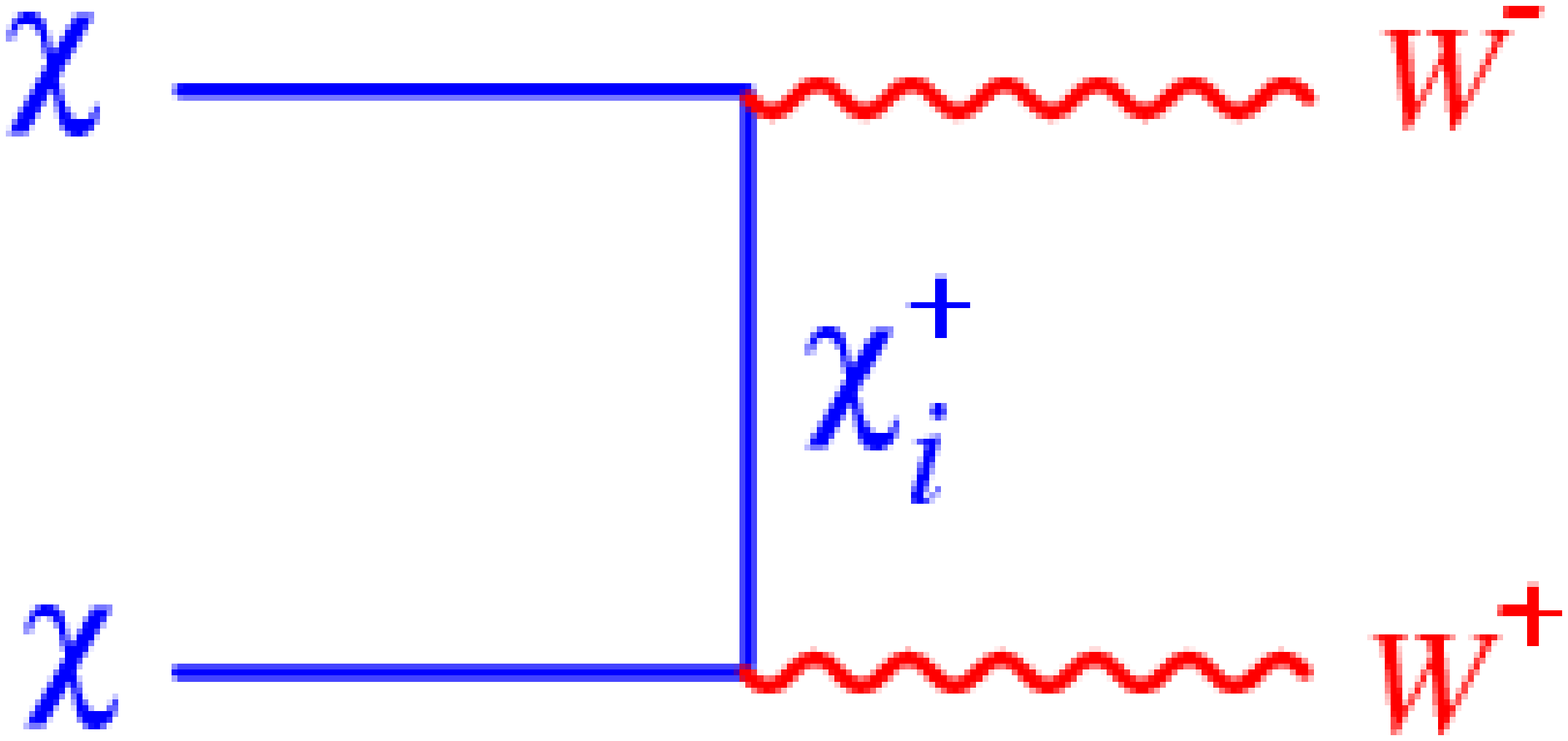}
\hspace*{0.2in}
\includegraphics[height=2.0in]{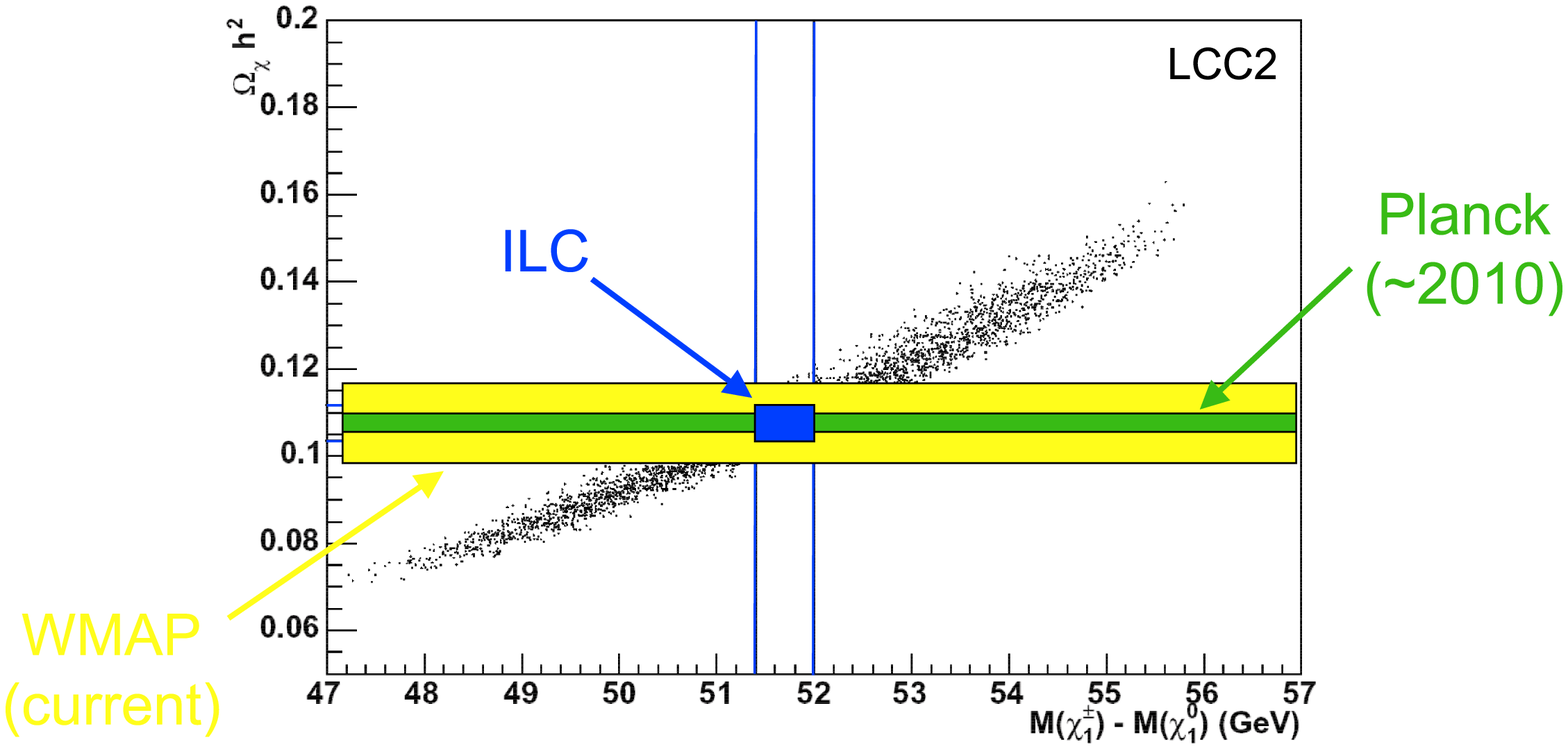}
\caption{Left: The dominant neutralino annihilation process in the
  focus point region.  Right: Constraints in the $(\mchi, \Omegachi)$
  plane from the ILC, with constraints on $\Omegachi$ from the WMAP
  and Planck satellite experiments.  The 50,000 scan points used to
  determine the ILC constraint are also shown (see
  text)~\cite{whitepaper}. Note that the distribution of scan points
  is much broader than the final ILC constrained region; out-lying
  points have very little probability weight. }
\label{fig:fp_feyn}
\end{figure*}

Determination of the thermal relic density in the focus point region
requires precise measurements of neutralino and chargino masses and
their mixings.  Applying the method described above for converting
collider constraints to a constraint on the thermal relic density, the
thermal relic density may be determined with fractional uncertainty
\begin{equation}
\text{LCC2 (preliminary):} \ 
\frac{\Delta (\Omegachi h^2)}{\Omegachi h^2} = 2.4\% 
\quad [ \Delta (\Omegachi h^2) = 0.0026 ] \ .
\end{equation}

\subsubsection{What We Learn}

The results of \figsref{bulk_omega}{fp_feyn} imply that the ILC will
provide a part per mille determination of $\Omegachi h^2$ in these
cases, matching WMAP and even the extraordinary precision expected
from Planck.  The many possible implications of such measurements are
outlined in the flowchart of \figref{flowchart}.

\begin{figure*}[t]
\centering
\includegraphics[width=0.95\textwidth]{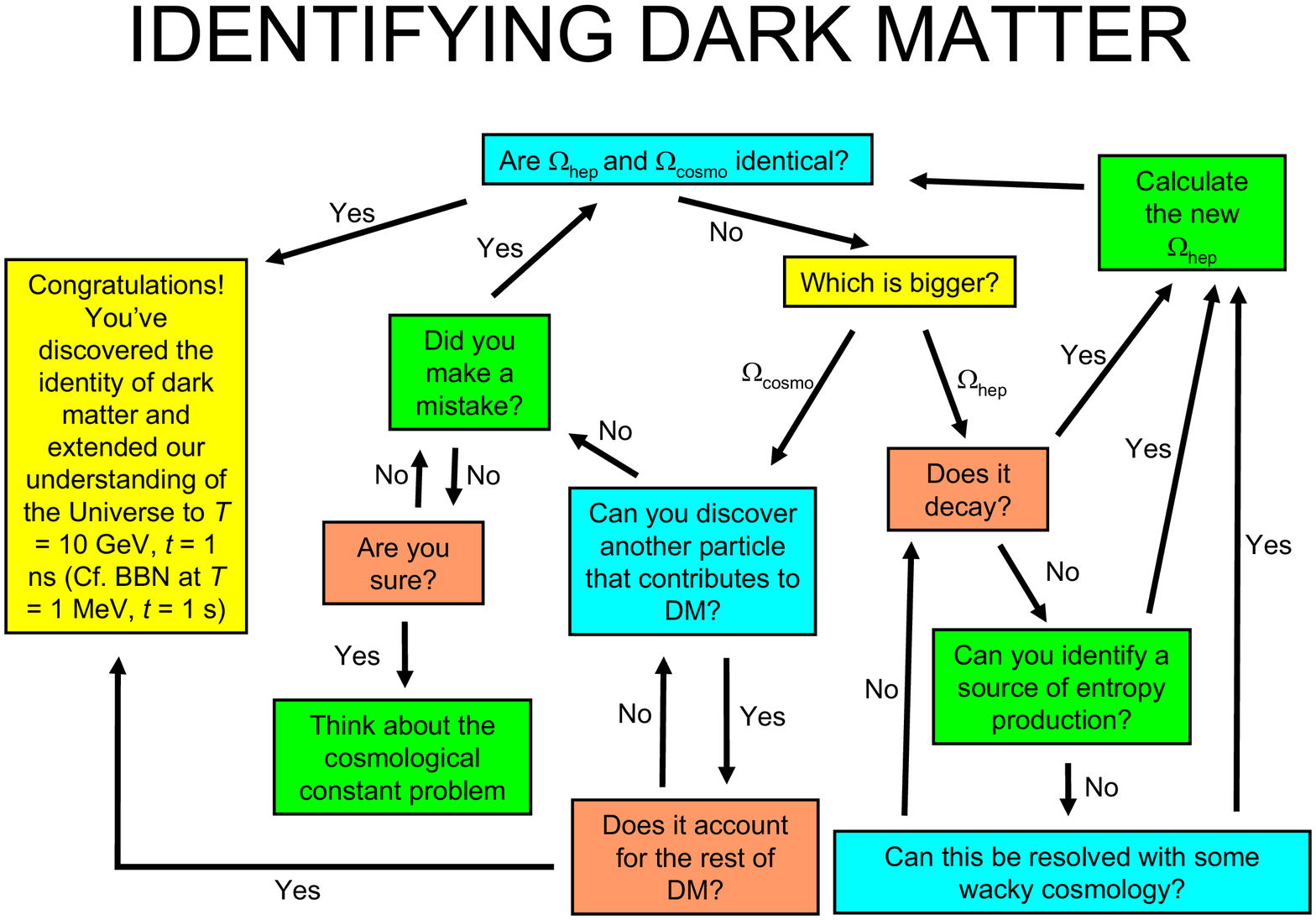}
\caption{Flowchart illustrating the possible implications of comparing
  $\Omega_{\text{hep}}$, the predicted dark matter thermal relic
  density determined from high energy physics, and
  $\Omega_{\text{cosmo}}$, the actual dark matter relic density
  determined by cosmological observations.}
\label{fig:flowchart}
\end{figure*}

Consistency of the ILC and WMAP/Planck measurements at the part per
mille level would provide strong evidence that neutralinos are
absolutely stable and form all of the non-baryonic dark matter.  Such
a result would at last provide convincing evidence that we have
produced dark matter at colliders and that we have identified its
microphysical properties.  It would be a landmark success of the
particle physics/cosmology connection, and would give us confidence in
our understanding of the Universe back to neutralino freeze out at
$t\sim 10^{-8}~\s$, eight orders of magnitude earlier than can
currently be claimed.

On the other hand, inconsistency would lead to a Pandora's box of
possibilities, all with important implications.  If the thermal relic
density determined from high energy physics is smaller than what is
required cosmologically, these high precision measurements imply that
neutralinos are at most only one component of cold, non-baryonic dark
matter.  On the other hand, if the thermal relic density determined at
colliders is too large, these measurements imply that neutralinos must
decay (perhaps to superWIMPs --- see below), or that the neutralino
thermal relic density is diluted by entropy production or some other
effect after freeze out.

The implications of LHC precision measurements for the relic density,
determined in the way discussed above, are also shown in
\figref{bulk_omega}.  The LHC precision in the LCC1 scenario is
extraordinary and unusual; for other scenarios, the LHC is unlikely to
determine $\Omegachi$ to better than one or more orders of magnitude.
At the same time, even in this ``best case scenario,'' the LHC
determination of $\Omegachi$ leaves open many possibilities.  For
example, comparison of the LHC result with WMAP/Planck cannot
differentiate between a Universe with only neutralino dark matter and
a Universe in which dark matter has two components, with neutralinos
making up only 80\%.  Such scenarios are qualitatively distinct, in
the sense that the possibility of another component with such
significant energy density can lead to highly varying conclusions
about the contents of the Universe and the evolution of structure that
formed the galaxies we see today.

\subsection{Mapping the WIMP Universe}

WIMPs may appear not only at colliders, but also in dark matter
searches.  Direct dark matter search experiments look for the recoil
of WIMPs scattering off highly shielded detectors.  Indirect dark
matter searches look for the products, such as positrons, gamma rays
or neutrinos, of WIMPs annihilating nearby, such as in the halo, the
galactic center, or the core of the Sun.

If WIMPs are discovered at colliders and their thermal relic densities
are determined to be cosmologically significant, it is quite likely
that they will also be discovered through direct and indirect dark
matter search experiments.  The requirement of the correct relic
density implies that WIMP annihilation was efficient in the early
Universe.  This suggests efficient annihilation now, corresponding to
significant indirect detection rates, and efficient scattering now,
corresponding to significant direct detection rates.  This rough
correspondence is illustrated in \figref{crossing}.

\begin{figure*}[t]
\centering
\includegraphics[height=2.0in]{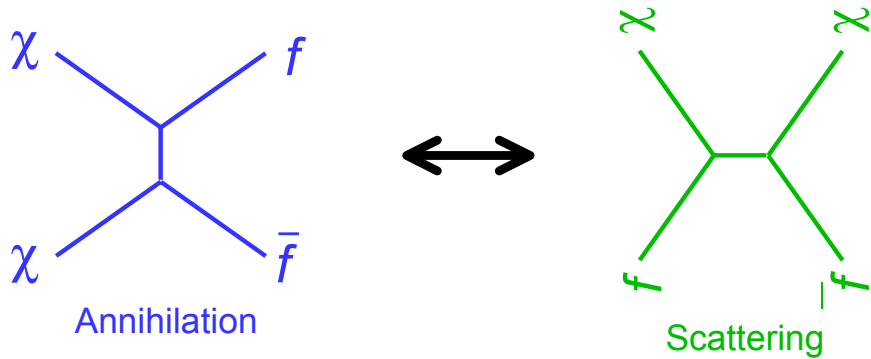}
\caption{Efficient annihilation, corresponding to large indirect
  detection rates, is related to efficient scattering, corresponding
  to large direct detection rates.}
\label{fig:crossing}
\end{figure*}

Direct and indirect dark matter detection rates are subject to
uncertainties from both particle physics, through the microphysical
properties of dark matter, and astrophysics, through the spatial and
velocity distributions of dark matter.  If completed, the research
program described in \secref{relic} to pin down the properties of
WIMPs will effectively remove particle physics uncertainties.  Dark
matter search experiments then become probes of dark matter
distributions.

As an example, consider direct detection.  Theoretical predictions of
direct detection rates are given in \figref{direct_detection}.  As is
typically done in particle physics studies, a simple dark matter halo
profile is assumed throughout this figure.  The enormous variation in
rates results from particle physics uncertainties alone.  LHC and ILC
studies will reduce this uncertainty drastically.  For example, for
LCC2, the dark matter mass will be determined to a GeV at the ILC, and
the cross section for neutralino-proton scattering will be determined
to $\Delta \sigma / \sigma \alt 10\%$~\cite{whitepaper}.  This
constraint is shown in \figref{direct_detection}, where the
uncertainties are smaller than the extent of the $\star$ plotting
symbol.

\begin{figure*}[t]
\centering
\includegraphics[width=0.95\textwidth]{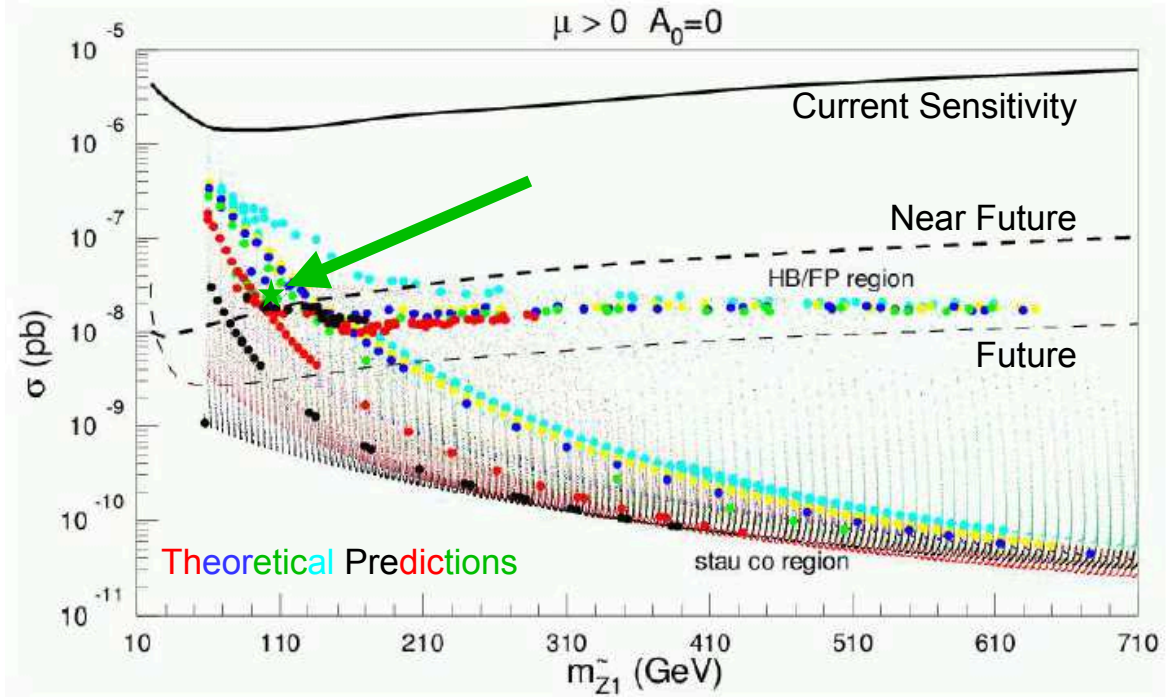}
\caption{Theoretical predictions for the direct detection
neutralino-proton scattering cross section $\sigma$, as a function of
neutralino mass $m_{\tilde{Z}_1} = \mchi$, for various mSUGRA models
(dots)~\cite{Baer:2003jb}, and the prediction of LCC2 ($\star$).  ILC
studies will constrain the values of $\sigma$ and $\mchi$ to be
smaller than the extent of the $\star$ plotting
symbol~\cite{whitepaper}.}
\label{fig:direct_detection}
\end{figure*}

Once collider constraints effectively remove microphysical
uncertainties, the direct detection rates give us information about
the local dark matter density and velocity profile.  In a similar way,
indirect detection rates will provide additional complementary
information.  For example, experiments such as HESS and GLAST may
detect photons from dark matter annihilation in the galactic center.
Such rates are sensitive to the halo profile at the galactic center, a
quantity of great interest at present.  The synergy between collider
experiments and these dark matter experiments will constrain the phase
space distribution of WIMP dark matter in the Universe.  Together with
$N$-body simulations, semi-analytical analyses of galaxy formation,
and cosmology observations, these data will have important
implications for the formation and evolution of structure.

\section{SuperWIMPs \label{sec:superwimps}}

In superWIMP scenarios~\cite{Feng:2003xh,Feng:2003uy}, a WIMP freezes
out as usual, but then decays to a stable dark matter particle that
interacts {\em superweakly}, as shown in \figref{freezeout_swimp}. The
prototypical example of a superWIMP is a weak-scale gravitino produced
non-thermally in the late decays of a weakly-interacting
next-to-lightest supersymmetric particle (NLSP), such as a neutralino,
charged slepton, or sneutrino~\cite{Feng:2003xh,%
Feng:2003uy,Ellis:2003dn,Feng:2004zu,Feng:2004mt,Wang:2004ib,%
Ellis:2004bx,Roszkowski:2004jd}.  Additional examples include
axinos~\cite{Rajagopal:1990yx,axinos} and
quintessinos~\cite{Bi:2003qa} in supersymmetry, Kaluza-Klein graviton
and axion states in models with universal extra
dimensions~\cite{Feng:2003nr}, and stable particles in models that
simultaneously address the problem of baryon
asymmetry~\cite{Kitano:2005ge}.  SuperWIMPs have all of the virtues of
WIMPs.  They exist in the same well-motivated frameworks and are
stable for the same reasons.  In addition, in many cases the WIMP and
superWIMP masses have the same origin.  In these cases, the decaying
WIMP and superWIMP naturally have comparable masses, and superWIMPs
also are automatically produced with relic densities of the desired
order of magnitude.

\begin{figure*}[t]
\centering \includegraphics[width=0.95\textwidth]{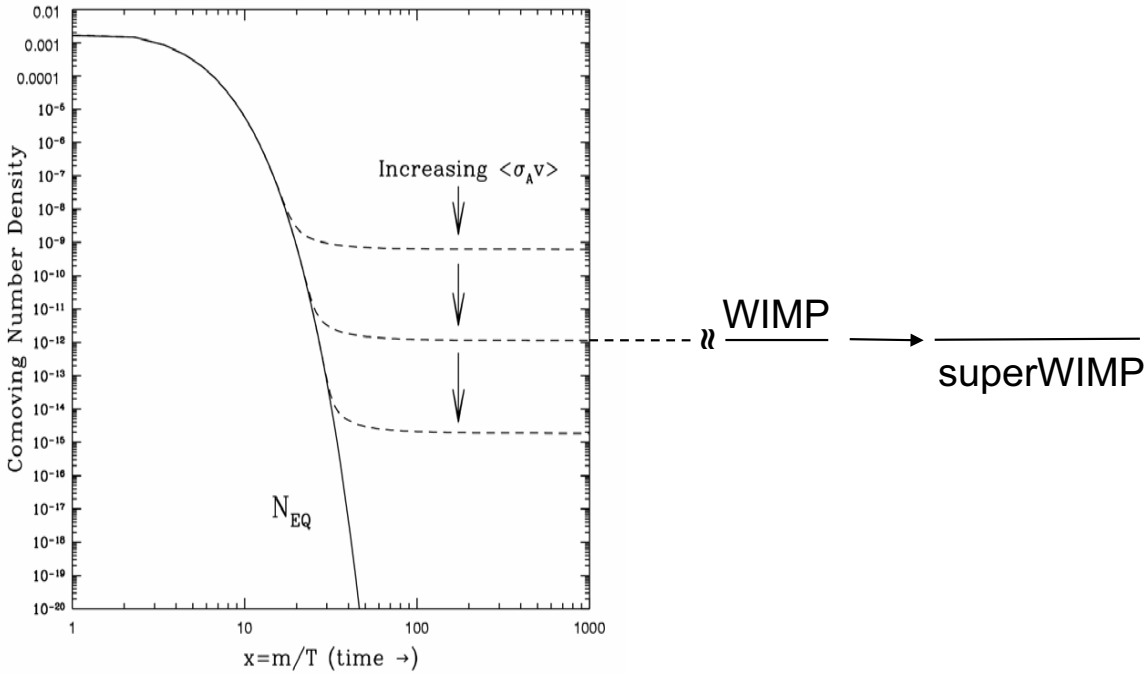}
\caption{In superWIMP scenarios, a WIMP freezes out as usual, but then
decays to a superWIMP, a superweakly-interacting particle that forms
dark matter.}
\label{fig:freezeout_swimp}
\end{figure*}

As noted above, superWIMPs exist in many different contexts.  We
concentrate here on the case of gravitino superWIMPs.  In the simplest
supersymmetric models, supersymmetry is transmitted to standard model
superpartners through gravitational interactions, and supersymmetry is
broken at a high scale.  The mass of the gravitino $\gravitino$ is
\begin{equation}
m_{\gravitino} = \frac{F}{\sqrt{3} \mstar} \ ,
\label{gravitinomass}
\end{equation}
and the masses of standard model superpartners are
\begin{equation}
\tilde{m} \sim \frac{F}{\mstar} \ ,
\label{tildem}
\end{equation}
where $\mstar = (8 \pi G_N)^{-1/2} \simeq 2.4 \times 10^{18}~\gev$ is
the reduced Planck scale and $F \sim (10^{11}~\gev)^2$ is the
supersymmetry breaking scale squared.  The precise ordering of masses
depends on unknown, presumably ${\cal O}(1)$, constants in
\eqref{tildem}.  As discussed in \secref{wimps}, most supergravity
studies assume that the LSP is a standard model superpartner, such as
the neutralino.

The gravitino may be the LSP, however.  In supergravity with
high-scale supersymmetry breaking, the gravitino has weak scale mass
$\mweak \sim 100~\gev$ and couplings suppressed by $\mstar$.  The
gravitino's extremely weak interactions imply that it is irrelevant
during thermal freeze out.  The NLSP therefore freezes out as usual,
and if the NLSP is a slepton, sneutrino, or neutralino, its thermal
relic density is again $\Omega_{\NLSP} \sim 0.1$.  However, eventually
the NLSP decays to its standard model partner and the gravitino.  The
resulting gravitino relic density is
\begin{equation}
\Omega_{\gravitino} = \frac{m_{\gravitino}}{\mNLSP} \Omega_{\NLSP} \ .
\label{swimpomega}
\end{equation}
In supergravity, where $m_{\gravitino} \sim \mNLSP$, the gravitino
therefore inherits a relic density of the right order to be much or
all of non-baryonic dark matter.  {\em The superWIMP gravitino
scenario preserves the prime virtue of WIMPs, namely that they give
the desired amount of dark matter without relying on the introduction
of new, fine-tuned energy scales.}

The superWIMP scenario differs markedly from other gravitino dark
matter scenarios~\cite{Pagels:ke,Weinberg:zq,Krauss:1983ik,%
Nanopoulos:1983up,Khlopov:pf,Ellis:1984eq,Ellis:1984er,%
Juszkiewicz:gg,Ellis:1990nb,Moroi:1993mb,Bolz:2000fu,%
Brandenburg:2004du}.  In the earliest proposals, gravitinos were
produced thermally at temperatures $T \sim \mplanck$, with
$\Omega_{\gravitino} \sim 0.1$ obtained by requiring $\mgravitino \sim
\kev$.  Such scenarios are disfavored now by the expectation of an
intervening era of inflation, which would dilute such a primordial
population.  After inflation, however, gravitinos may be produced in
an era of reheating.  In this case, $\Omega_{\gravitino} \sim 0.1$ is
obtained for reheat temperatures $T_{\text{RH}} \sim 10^{10}~\gev$.
In contrast to the superWIMP scenario, it is not clear that gravitino
production during reheating has testable consequences, other than the
existence of cold dark matter itself.  In addition, the reheating
scenario requires the introduction of a new scale, in contrast to the
superWIMP production mechanism, where the relic density is a function
of the Fermi and Planck scales only and is naturally in the desired
range.  It is important to note, however, that the reheating and
superWIMP production mechanisms are not mutually exclusive.  The
current gravitino relic population may have components from both
production mechanisms, resulting in a very simple scenario in which
dark matter is composed of two populations of particles with different
histories and effects on the early Universe~\cite{Cerdeno:2005eu}.

Because superWIMP gravitinos interact only gravitationally, with
couplings suppressed by $\mstar$, they are impossible to detect in
conventional direct and indirect dark matter search experiments.  At
the same time, the extraordinarily weak couplings of superWIMPs imply
other testable signals.  The NLSP is a weak-scale particle decaying
gravitationally and so has a natural lifetime of
\begin{equation}
\frac{\mstar^2}{\mweak^3} \sim 10^4 - 10^8~\s \ .
\label{lifetime}
\end{equation}
This decay time, outlandishly long by particle physics standards,
implies testable cosmological signals, as well as novel signatures at
colliders.

\subsection{Cosmology}
\label{sec:cosmology}

The most sensitive probes of late decays with lifetimes in the range
given in \eqref{lifetime} are from Big Bang nucleosynthesis (BBN) and
the Planckian spectrum of the cosmic microwave background (CMB).  The
impact of late decays to gravitinos on BBN and the CMB are determined
by only two parameters: the lifetime of NLSP decays and the energy
released in these decays.  The energy released is quickly thermalized,
and so the cosmological signals are insensitive to the details of the
energy spectrum and are determined essentially only by the total
energy released.

The width for the decay of a slepton to a gravitino is
\begin{equation}
 \Gamma(\tilde{l} \to l \tilde{G}) =\frac{1}{48\pi \mstar^2}
 \frac{m_{\tilde{l}}^5}{m_{\tilde{G}}^2} 
 \left[1 -\frac{m_{\tilde{G}}^2}{m_{\tilde{l}}^2} \right]^4 \ ,
\label{sfermionwidth}
\end{equation}
assuming the lepton mass is negligible. (Similar expressions hold for
the decays of neutralino NLSPs.)  This decay width depends on only the
slepton mass, the gravitino mass, and the Planck mass.  In many
supersymmetric decays, dynamics brings a dependence on many
supersymmetry parameters.  In contrast, as decays to the gravitino are
gravitational, dynamics is determined by masses, and so no additional
parameters enter.  In particular, there is no dependence on left-right
mixing or flavor mixing in the slepton sector.  For $\mgravitino /
m_{\tilde{l}} \approx 1$, the slepton decay lifetime is
\begin{eqnarray}
 \tau(\tilde{l} \to l \tilde{G})
\simeq 3.6\times 10^8~\s
\left[\frac{100~\gev}{m_{\tilde{l}} - m_{\gravitino}}\right]^4
\left[\frac{m_{\tilde{G}}}{\tev}\right]\ .
\label{eq:decaylifetime}
\end{eqnarray}
This expression is valid only when the gravitino and slepton are
nearly degenerate, but it is a useful guide and verifies the rough
estimate of \eqref{lifetime}.

The energy release is conveniently expressed in terms of
\begin{eqnarray}
\xi_{\text{EM}} \equiv \epsilon_{\text{EM}} B_{\text{EM}}
Y_{\text{NLSP}}
\label{eq:xi_EM}
\end{eqnarray}
for electromagnetic energy, with a similar expression for hadronic
energy.  Here $\epsilon_{\text{EM}}$ is the initial EM energy released
in NLSP decay, and $B_{\text{EM}}$ is the branching fraction of NLSP
decay into EM components.  $Y_{\text{NLSP}} \equiv
n_{\text{NLSP}}/n_{\gamma}$ is the NLSP number density just before
NLSP decay, normalized to the background photon number density
$n_{\gamma} = 2 \zeta(3) T^3 / \pi^2$.  It can be expressed in terms
of the superWIMP abundance:
\begin{equation}
Y_{\text{NLSP}}\simeq 3.0 \times 10^{-12}
\left[\frac{\tev}{m_{\gravitino}}\right]
\left[\frac{\Omega_{\gravitino}}{0.23}\right] \ .
\label{eq:def_Y}
\end{equation}

Once an NLSP candidate is specified, and assuming superWIMPs make up
all of the dark matter, with $\Omega_{\gravitino} = \Omega_{\text{DM}}
= 0.23$, the early Universe signals are completely determined by only
two parameters: $m_{\gravitino}$ and $m_{\NLSP}$.

\subsubsection{BBN Electromagnetic Constraints}

BBN predicts primordial light element abundances in terms of one free
parameter, the baryon-to-photon ratio $\eta \equiv n_B / n_{\gamma}$.
In the past, the fact that the observed D, $^4$He, $^3$He, and $^7$Li
abundances could be accommodated by a single choice of $\eta$ was a
well-known triumph of standard Big Bang cosmology.

More recently, BBN baryometry has been supplemented by CMB data, which
alone yields $\eta_{10} = \eta / 10^{-10} = 6.1 \pm
0.4$~\cite{Spergel:2003cb}.  This value agrees precisely with the
value of $\eta$ determined by D, considered by many to be the most
reliable BBN baryometer.  However, it highlights slight
inconsistencies in the BBN data.  Most striking is the case of $^7$Li.
For $\eta_{10} = 6.0\pm 0.5$, the value favored by the combined D and
CMB observations, the standard BBN prediction is~\cite{Burles:2000zk}
\begin{eqnarray}
^7\text{Li/H} &=& 4.7_{-0.8}^{+0.9} \times 10^{-10} \label{Li}
\end{eqnarray}
at 95\% CL.  This contrasts with observations.
Three independent studies find 
\begin{eqnarray}
\text{$^7$Li/H} &=& 1.5_{-0.5}^{+0.9} \times 10^{-10} \quad 
\text{(95\% CL)~\cite{Thorburn}} \\
\text{$^7$Li/H} &=& 1.72_{-0.22}^{+0.28} \times 10^{-10} \ 
\text{($1\sigma + \text{sys}$)~\cite{Bonafacio}} \\
\text{$^7$Li/H} &=& 1.23_{-0.32}^{+0.68} \times 10^{-10} \ 
\text{(stat + sys, 95\% CL)~\cite{Ryan:1999vr}} \ ,
\end{eqnarray}
where depletion effects have been estimated and included in the last
value.  Within the published uncertainties, the observations are
consistent with each other but inconsistent with the theoretical
prediction of \eqref{Li}, with central values lower than predicted by
a factor of 3 to 4.  $^7$Li may be depleted from its primordial value
by astrophysical effects, for example, by rotational mixing in stars
that brings Lithium to the core where it may be
burned~\cite{Pinsonneault:1998nf,Vauclair:1998it}, but it is
controversial whether this effect is large enough and consistent with
the relatively small scatter of observations to reconcile observations
with the BBN prediction~\cite{Ryan:1999vr}.

We now consider the effects of NLSP decays to gravitinos.  For
weakly-interacting NLSPs, that is, sleptons, sneutrinos, and
neutralinos, the energy released is dominantly deposited in
electromagnetic cascades.  For the decay times of \eqref{lifetime},
mesons decay before they interact hadronically.  The impact of EM
energy on the light element abundances has been studied in
Refs.~\cite{Kawasaki:1994sc,Holtmann:1998gd,%
Kawasaki:2000qr,Cyburt:2002uv,Ellis:2005ii}.  The results of
Ref.~\cite{Cyburt:2002uv} are given in \figref{stau}.  The shaded
regions are excluded because they distort the light element abundances
too much.  The predictions of the superWIMP scenario for a stau NLSP
with $m_{\gravitino}$ and $\m_{\NLSP}$ varying over weak scale
parameters are given in \figref{stau} by the grid.

\begin{figure}[tb]
\begin{center}
\includegraphics*[width=11cm]{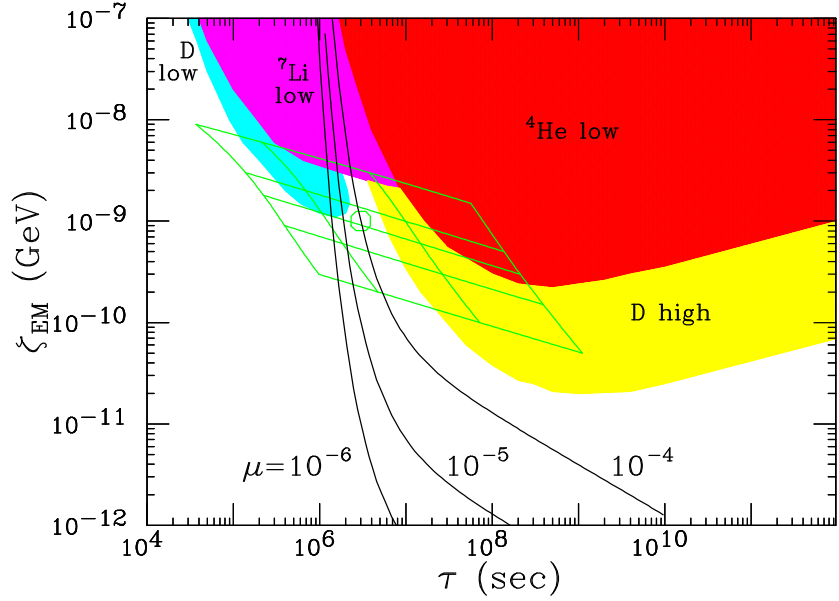}
\caption{Predicted and excluded regions of the $(\tau,
\zeta_{\text{EM}})$ plane in the superWIMP dark matter scenario, where
$\tau$ is the lifetime for $\tilde{l} \to l \tilde{G}$, and
$\zeta_{\text{EM}}$ is the normalized electromagnetic energy release.
The grid gives predicted values for $m_{\tilde{G}} = 100~\gev -
3~\tev$ (top to bottom) and $\Delta m \equiv m_{\tilde{l}} -
m_{\tilde{G}} = 600~\gev - 100~\gev$ (left to right), assuming
$\Omega_{\tilde{G}} = 0.23$.  BBN constraints exclude the shaded
regions; the circle indicates the best fit region where $^7$Li is
reduced to observed levels without upsetting other light element
abundances.  Contours of CMB $\mu$ distortions indicate the current
bound ($\mu < 0.9 \times 10^{-4}$) and the expected future sensitivity
of Diffuse Microwave Emission Survey (DIMES) ($\mu \sim 10^{-6}$).
{}From Ref.~\protect\cite{Feng:2003uy}.  }
\label{fig:stau}
\end{center}
\end{figure}

We find that the BBN constraint excludes some weak scale parameters.
However, much of the weak scale parameter space remains viable.  Note
also that, given the $^7$Li discrepancy, the best fit is not achieved
at $\xi_{\text{EM}} = 0$, but rather for $\tau \sim 3\times 10^6~\s$
and $\xi_{\text{EM}} \sim 10^{-9}~\gev$, where
$^7$Li is destroyed by late decays without
changing the other relic abundances.  This point is marked by the
circle in \figref{stau}.  The energy release predicted in the
superWIMP scenario naturally includes this region.  The $^7$Li anomaly
is naturally resolved in the superWIMP scenario by a stau NLSP with
$m_{\NLSP} \sim 700~\gev$ and $m_{\gravitino} \sim 500~\gev$.

\subsubsection{BBN Hadronic Constraints}

Hadronic energy release is also constrained by BBN~\cite{Reno:1987qw,%
Dimopoulos:1987fz,Dimopoulos:1988ue,Kohri:2001jx,Jedamzik:2004er,%
Kawasaki:2004yh}.  In fact, constraints on hadronic energy release are
so severe that even subdominant contributions to hadronic energy may
provide stringent constraints.

Slepton and sneutrino decays contribute to hadronic energy through the
higher order processes
\begin{eqnarray}
\tilde{l} &\to& l Z \tilde{G} \ , \ \nu W \tilde{G} \nonumber \\
\tilde{\nu} &\to& \nu Z \tilde{G} \ , \ l W \tilde{G} \ ,
\end{eqnarray}
when the $Z$ or $W$ decays hadronically.  These three-body decays may
be kinematically suppressed when $m_{\tilde{l}, \tilde{\nu}} -
m_{\tilde{G}} < m_W, m_Z$, but even in this case, four-body decays,
such as $\tilde{l} \to l \gamma^* \tilde{G} \to l q\bar{q} \tilde{G}$,
contribute to hadronic cascades and may be important.  The branching
fractions for these decays have been calculated in
Refs.~\cite{Feng:2004zu,Feng:2004mt}.  The end result is that these
constraints are stringent and important, as they exclude regions of
parameter space that would otherwise be allowed.  At the same time,
much of the parameter space in the case of slepton and sneutrino NLSPs
remains viable.  For details, see
Refs.~\cite{Feng:2004zu,Feng:2004mt}.

In contrast to the case of slepton and sneutrino NLSPs, the neutralino
NLSP possibility is very severely constrained by bounds on hadronic
energy release.  This is because neutralinos contribute to hadronic
energy even through two-body decays 
\begin{equation}
\chi \to Z \gravitino , \ h \gravitino \ ,
\end{equation}
followed by $Z,h \to q \bar{q}$.  The resulting hadronic cascades
destroy BBN successes, and exclude this scenario unless such decays
are highly suppressed.  Kinematic suppression is not viable, however
--- if $m_{\chi} - m_{\gravitino} < m_Z$, the decay $\chi \to \gamma
\gravitino$ takes place so late that it violates bounds on EM
cascades.  {\em Neutralino NLSPs are therefore highly
disfavored}~\cite{Feng:2003xh,Feng:2003uy,Feng:2004zu,%
Feng:2004mt,Roszkowski:2004jd}; they are allowed only when the
two-body decays to $Z$ and $h$ bosons are suppressed dynamically, as
when the neutralino is photino-like, a possibility that is not
well-motivated by high energy frameworks.

\subsubsection{CMB Constraints}

The injection of electromagnetic energy may also distort the frequency
dependence of the CMB black body
radiation~\cite{Hu:1993gc,Lamon:2005jc}.  For the decay times of
interest, with redshifts $z \sim 10^5$ to $10^7$, the resulting
photons interact efficiently through $\gamma e^- \to \gamma e^-$ and
$e X \to e X \gamma$, where $X$ is an ion, but photon number is
conserved, since double Compton scattering $\gamma e^- \to \gamma
\gamma e^-$ is inefficient.  The spectrum therefore relaxes to
statistical but not thermodynamic equilibrium, resulting in a
Bose-Einstein distribution function
\begin{equation}
f_{\gamma}(E) = \frac{1}{e^{E/(kT) + \mu} - 1} \ ,
\end{equation}
with chemical potential $\mu \ne 0$.

In \figref{stau} we show contours of chemical potential $\mu$, as
determined by updating the analysis of Ref.~\cite{Hu:1993gc}.  (For a
more recent analysis and its implications for superWIMPs, see
Ref.~\cite{Lamon:2005jc}.)  The current bound is $\mu < 9\times
10^{-5}$~\cite{Fixsen:1996nj,Eidelman:2004wy}. We see that, although
there are at present no indications of deviations from black body,
current limits are already sensitive to the superWIMP scenario, and
are even beginning to probe regions favored by the BBN considerations
described above. In the future, the Absolute Radiometer for Cosmology,
Astrophysics, and Diffuse Emission (ARCADE) and Diffuse Microwave
Emission Survey (DIMES) experiments may improve sensitivities to $\mu
\approx 2 \times 10^{-6}$~\cite{ARCADE}.  ARCADE and DIMES will
therefore probe further into superWIMP parameter space, and will
effectively probe all of the favored region where the $^7$Li
underabundance is explained by decays to superWIMPs.

\subsection{Colliders}
\label{sec:colliders}

The study of superWIMP dark matter at colliders has elements in common
with the study of WIMPs, but with key differences.  It may also be
divided into three (overlapping) stages:
\begin{enumerate}
\item SuperWIMP Candidate Identification.  Is there evidence for late
  decays to superWIMPs from collider studies?
\item SuperWIMP Relic Density Determination. What are the superWIMP
  candidates' predicted relic densities?  Can they be significant
  components or all of dark matter?  What are their masses, spins, and
  other quantum numbers?
\item Mapping the SuperWIMP Universe. Combined with other
  astrophysical and cosmological results, what can collider studies
  tell us about astrophysical questions, such as the distribution of
  dark matter in the Universe?
\end{enumerate}

For Stage 1, collider evidence for superWIMPs may come in one of two
forms.  Collider experiments may find evidence for charged, long-lived
particles.  Given the stringent bounds on charged dark matter, such
particles presumably decay, and their decay products may be
superWIMPs.  Alternatively, colliders may find seemingly stable WIMPs,
but the WIMP relic density studies described in \secref{relic} may
favor a relic density that is too large, a conundrum that may be
resolved by postulating that WIMPs decay.  These two possibilities are
not mutually exclusive.  In fact, the discovery of charged long-lived
particles with too-large predicted relic density is a distinct
possibility and would provide strong motivation for superWIMP dark
matter.

In the following subsections, we will explore how well the LHC and ILC
may advance Stages 2 and 3.

\subsection{Relic Density Determination}

SuperWIMPs are produced in the late decays of WIMPs.  Their number
density is therefore identical to the WIMP number density at freeze
out, and so, as noted in \eqref{swimpomega}, the superWIMP relic
density is
\begin{equation}
\Omega_{\text{sWIMP}} = \frac{m_{\text{sWIMP}}}{m_{\text{WIMP}}}
\Omega_{\text{WIMP}} \ .
\label{swimp_omega}
\end{equation}
To determine the superWIMP relic density, we must therefore determine
the superWIMP's mass.  This is not easy, since the WIMP lifetime may
be very large, implying that superWIMPs are typically produced long
after the WIMPs have escaped collider detectors.

For concreteness, consider the case of supersymmetry with a stau NLSP
decaying to a gravitino superWIMP. (Recall that, if superWIMPs are
produced in sufficient numbers to be much of the dark matter,
neutralino NLSPs are heavily disfavored, as their late decays
invariably violate constraints from BBN and the
CMB~\cite{Feng:2003xh,Feng:2003uy,Feng:2004zu,%
Feng:2004mt,Roszkowski:2004jd}.)  As discussed above, the stau's
lifetime is outlandishly long by particle physics standards.  This
gravitino superWIMP scenario therefore implies that the signal of
supersymmetry at colliders will be meta-stable sleptons with lifetimes
of days to months.  Given their large mass, some of these sleptons
will be slow, and so will produce highly-ionizing tracks that should
be spectacularly obvious at the
LHC~\cite{Drees:1990yw,Goity:1993ih,Nisati:1997gb,Feng:1997zr}.

At the same time, because some sleptons will be slowly moving and
highly-ionizing, they may be trapped and
studied~\cite{Feng:2004yi,Hamaguchi:2004df,Brandenburg:2005he,%
DeRoeck:2005bw}.  As an example, sleptons may be trapped in water
tanks placed outside collider detectors.  These water tanks may then
be drained periodically to underground reservoirs where slepton decays
may be observed in quiet environments.  This possibility has been
studied in Ref.~\cite{Feng:2004yi} and is illustrated in
\figref{trap_ILC}.  Alternatively, sleptons may be trapped in the
detectors themselves~\cite{Feng:2004yi,Hamaguchi:2004df}, or may be
stopped in the surrounding rock~\cite{DeRoeck:2005bw}.

\begin{figure*}[t]
\centering
\includegraphics[height=2.8in]{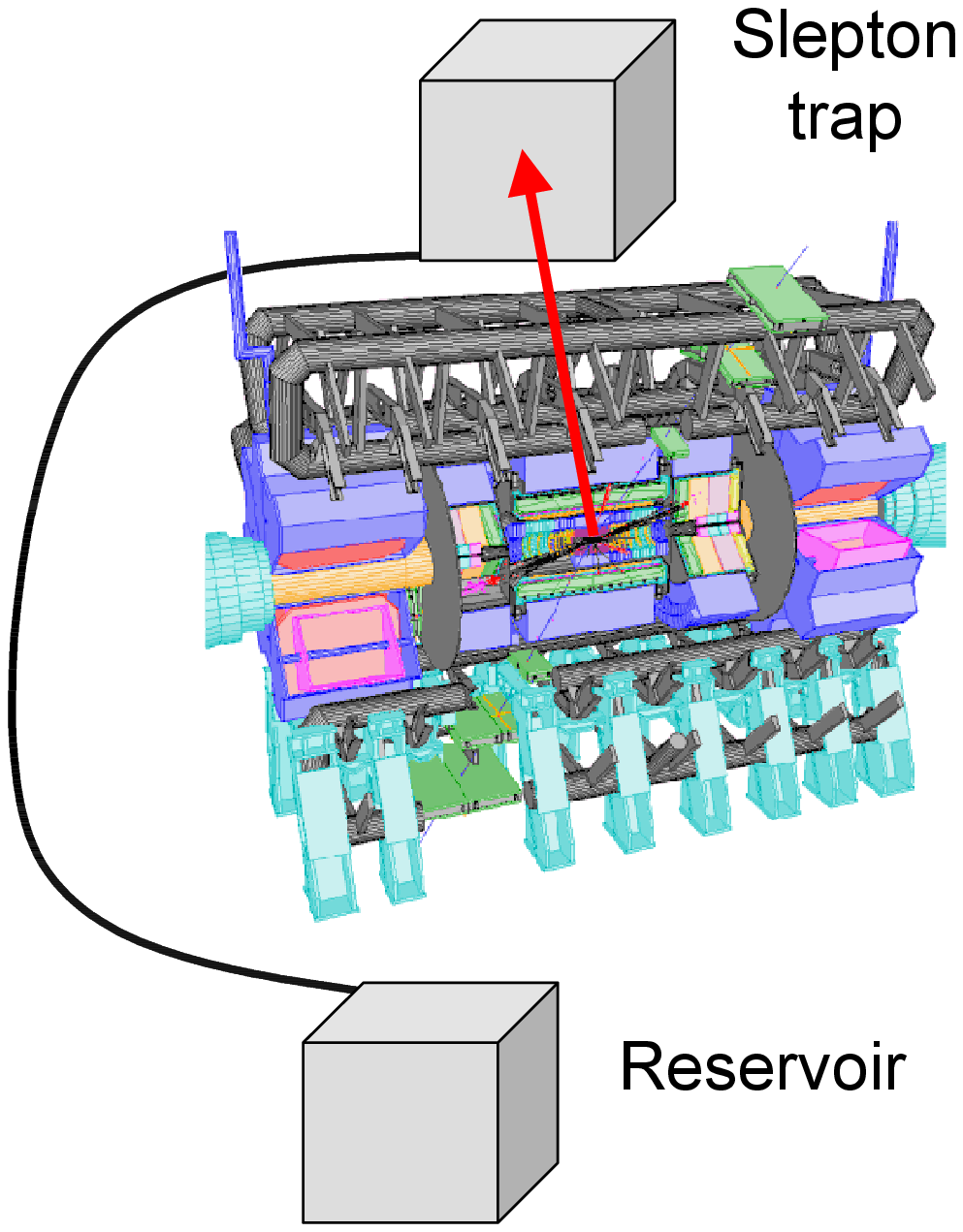}
\hspace*{0.2in}
\includegraphics[height=2.8in]{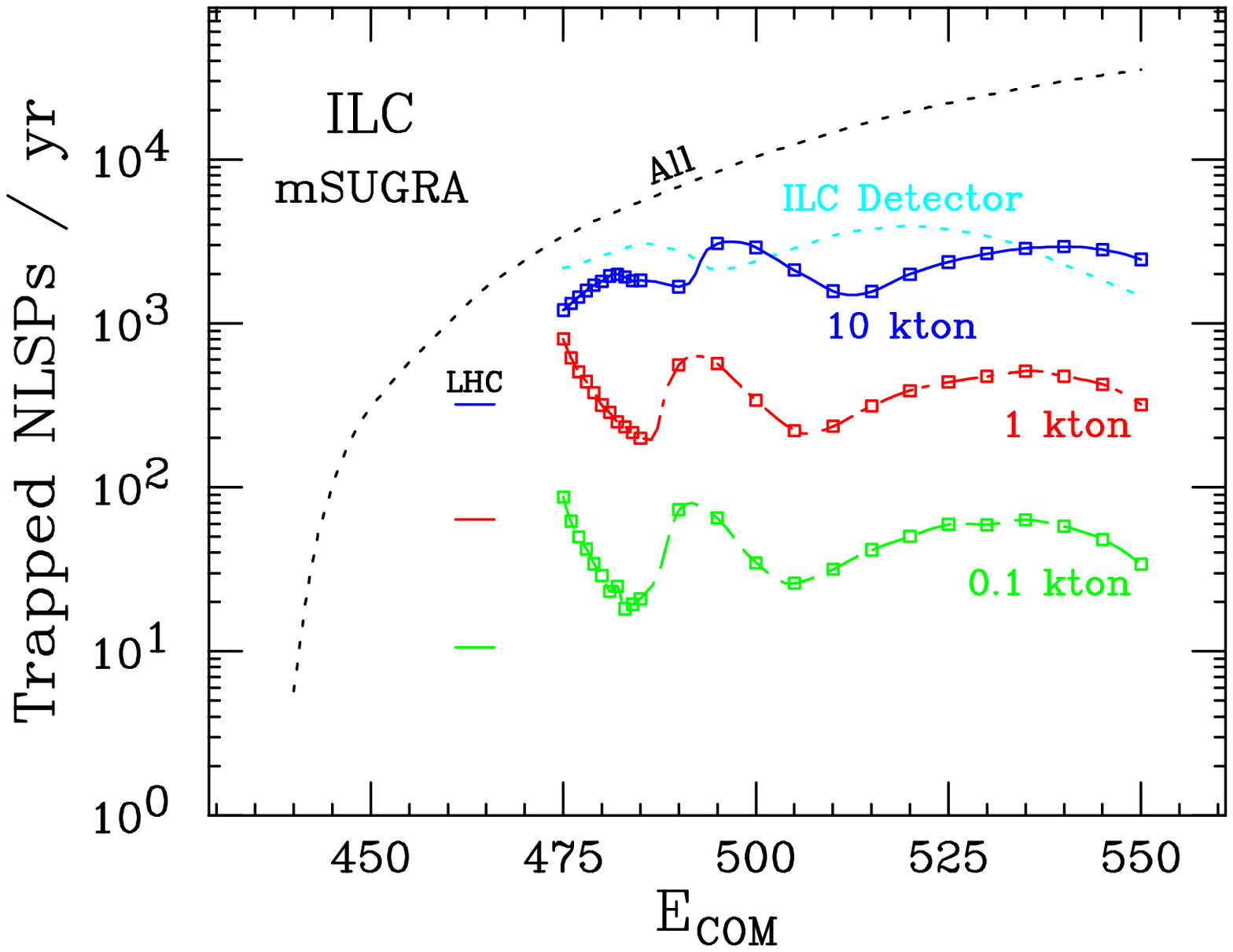}
\caption{Left: Configuration for slepton trapping in gravitino
superWIMP scenarios.  Right: The number of sleptons trapped per year
at the ILC in 10 kton (solid), 1 kton (dot-dashed), and 0.1 kton
(dashed) water traps.  The total number of sleptons produced is also
shown (upper dotted) along with the number of sleptons trapped in the
ILC detector (lower dotted).  The trap shape and placement have been
optimized and a luminosity of $300~\ifb/\yr$ is assumed.  The
underlying model is minimal supergravity with $\mgaugino = 600~\gev$,
$m_0 = 0$, $A_0 = 0$, $\tan\beta = 10$, and $\mu > 0$.  The LHC
results for this model are as indicated~\protect\cite{Feng:2004yi}.}
\label{fig:trap_ILC}
\end{figure*}

How many sleptons may be stopped in a reasonably sized trap?  The
answer is highly model-dependent.  The results for one model with 219
GeV sleptons is shown in \figref{trap_ILC}.  At the LHC, hundreds of
sleptons may be caught each year in a 10 kton trap, assuming a
luminosity of $100~\ifb/\yr$.  A 10 kton trap is not particularly
bulky. The optimal shape is one that covers as much solid angle as
possible and is only $\sim 1~\m$ thick~\cite{Feng:2004yi}.  These LHC
results may be improved significantly if long-lived NLSP sleptons are
kinematically accessible at the ILC.  For the identical case with 219
GeV sleptons, ${\cal O}(1000)$ sleptons may be trapped each year in a
10 kton trap at the ILC, assuming $300~\ifb/\yr$.  By considering the
slightly more general possibility of placing lead or other dense
material between the ILC detector and the slepton trap, a further
enhancement of an order of magnitude may be possible, allowing up to
${\cal O}(10^4)$ sleptons to be trapped per ILC year.  These ILC
results are made possible by the ability to tune the beam energy to
produce slow NLSPs.  The ability to prepare initial states with
well-known energies and the flexibility to tune this energy are
well-known advantages of the ILC.  Here, these features are exploited
in a qualitatively new way to produce slow sleptons that are easily
captured.

If thousands of sleptons are trapped, the slepton lifetime may be
determined to the few percent level simply by counting the number of
slepton decays as a function of time.  The slepton mass will be
constrained by analysis of the collider event kinematics. A percent
level measurement of the slepton lifetime given in
\eqref{sfermionwidth} therefore implies a high precision measurement
of the gravitino mass, and therefore a determination of the gravitino
relic density through \eqref{swimp_omega}.  As with the case of WIMPs,
consistency at the percent level with the observed dark matter relic
density will provide strong evidence that dark matter is indeed
composed of gravitino superWIMPs.

SuperWIMP quantum numbers and couplings may also be determined through
collider studies~\cite{Buchmuller:2004rq,Feng:2004gn}, although, as
indicated above, these will typically be determined after or at the
same time as the relic density determination, in contrast to the case
of WIMPs.  For example, an alternative method to determine the
gravitino mass is to measure the energy of slepton decay products.
This provides a consistency check of the mass determination described
above.  Alternatively, these two methods, when combined, determine not
only $m_{\gravitino}$, but also the Planck mass $\mstar$.  Given
enough events, the gravitino spin may also be constrained to be 3/2
through detailed analyses of angular
distributions~\cite{Buchmuller:2004rq}, and the gravitino
differentiated from other superWIMP candidates, such as the
axino~\cite{Brandenburg:2005he}.  The spin and couplings of the
gravitino may therefore be determined, showing that the superWIMP is
in fact the superpartner of the graviton and that nature is locally
supersymmetric.

\subsection{Mapping the SuperWIMP Universe}

Collider studies of superWIMPs will have significant implications for
the phase space distribution of dark matter.  In fact, the discovery
of superWIMPs may resolve current discrepancies and shed light on
important and controversial issues in structure formation.

In the standard cosmology, dark matter is assumed to be cold, as is
the case with WIMP dark matter.  Cold dark matter is remarkably
successful in explaining the observed large scale structure down to
length scales of $\sim 1~\Mpc$.  Despite its considerable virtues,
however, cold dark matter appears to face difficulty in explaining the
observed structure on length scales $\alt 1~\Mpc$. Numerical
simulations assuming cold dark matter predict, for example, overdense
cores in galactic halos~\cite{Moore:1994yx} and too many dwarf
galaxies in the Local Group~\cite{Klypin:1999uc}.

These problems may be alleviated or resolved by superWIMP dark
matter~\cite{Sigurdson:2003vy,Profumo:2004qt,Kaplinghat:2005sy,%
Cembranos:2005us,Jedamzik:2005sx}.  SuperWIMPs are produced with
relativistic velocities at late times, as we have seen.  They
therefore exhibit properties typically associated with warm dark
matter, suppressing power on small scales and potentially resolving
the problems of cold dark matter mentioned above.\footnote{Note that
gravitino's produced thermally during reheating are cold, and do not
differ from standard cold dark matter in their impact in structure
formation.}  The discovery of superWIMP dark matter and the
determination of NLSP and superWIMP masses and other relevant
parameters at colliders would therefore change fundamentally our
understanding of how galaxies were formed and provide a new framework
for understanding halo profiles and the distribution of dark matter.

As discussed above, decays that produce superWIMPs also typically
release electromagnetic and hadronic energy.  This energy may modify
the light element abundances predicted by standard
BBN~\cite{Feng:2003xh,Feng:2003uy,%
Cyburt:2002uv,Jedamzik:2004er,Kawasaki:2004yh,Ellis:2005ii} or distort
the black body spectrum of the
CMB~\cite{Feng:2003xh,Feng:2003uy,Hu:1993gc,Fixsen:1996nj,Lamon:2005jc}.
Collider studies will be able to determine how much energy is released
and at what time, providing still more information with important
consequences for astrophysics and cosmology.

\section{Conclusions \label{sec:conclusions}}

This is an exciting time at the boundary of particle physics and
cosmology.  While important microphysical questions related to
electroweak symmetry breaking and flavor remain, breakthroughs in
cosmology have added a whole new layer of fundamental problems
requiring particle physics answers.

Many of the key problems revolve around the mysteries of dark matter.
Although there are many viable proposals, we have considered
particularly well-motivated candidates whose relic densities are tied
to two known energy scales, the Fermi and Planck scales, and fall
``coincidentally'' in the desired range.  There are two classes of
dark matter candidates with this property: WIMPs and superWIMPs.  In
both of these scenarios, dark matter particles, and typically many
others, are expected with masses at the Fermi scale $M_F \sim
100~\gev$.  In the next few years, particle colliders will at last
probe this scale.  If there is new physics, experiments at the
Tevatron, LHC, and ILC will likely discover it and study it in great
detail.

If dark matter is composed of WIMPs, the LHC, and particularly the
proposed ILC will be able to determine the WIMP's properties, and may
also pin down its thermal relic density.  If these determinations match
cosmological observations to high precision, then (and only then) we
will be able to claim to have determined what dark matter is. Such an
achievement will also lead, through synergy with direct and indirect
dark matter searches and cosmological observations, to improved
knowledge of dark matter distributions and the formation of large
scale structure.

If dark matter is composed of superWIMPs, the LHC and ILC will again
play a crucial role.  The signal of new physics will likely be
long-lived charged particles, a spectacular signature that may be
evident even in the first year or two of LHC running.  By trapping
these metastable particles and watching them decay, the properties and
relic density of superWIMPs may also be determined, providing another
opportunity to identify dark matter.  In the superWIMP scenario, the
identification of ``warm'' dark matter may also resolve current issues
in structure formation, and will also have implications for Big Bang
nucleosynthesis and the cosmic microwave background.

If any of the ideas discussed here is realized in nature, the coming
years of exploration will not only provide our first incisive look at
the Fermi scale, but will also yield profound insights about the
Universe, its contents, and its evolution.

\begin{acknowledgments}
For many contributions to the viewpoints and results summarized here,
it is a pleasure to thank James Bullock, Manoj Kaplinghat, Alan Kogut,
Ned Wright and the members of the ALCPG Cosmology Subgroup,
particularly my co-editors Marco Battaglia, Norman Graf, Michael
Peskin, and Mark Trodden and collaborators Jose Ruiz Cembranos,
Konstantin Matchev, Arvind Rajaraman, Bryan Smith, Shufang Su,
Fumihiro Takayama, and Frank Wilczek.  The work of JLF is supported in
part by NSF CAREER grant No.~PHY--0239817, NASA Grant No.~NNG05GG44G,
and the Alfred P.~Sloan Foundation.
\end{acknowledgments}


\begin{thebibliography}{999}

\bibitem{Knop:2003iy}
  R.~A.~Knop {\it et al.}  [The Supernova Cosmology Project Collaboration],
  Astrophys.\ J.\  {\bf 598}, 102 (2003)
  [astro-ph/0309368].

\bibitem{Spergel:2003cb}
  D.~N.~Spergel {\it et al.}  [WMAP Collaboration],
  Astrophys.\ J.\ Suppl.\  {\bf 148}, 175 (2003)
  [astro-ph/0302209].

\bibitem{Tegmark:2003ud}
  M.~Tegmark {\it et al.}  [SDSS Collaboration],
  Phys.\ Rev.\ D {\bf 69}, 103501 (2004)
  [astro-ph/0310723].

\bibitem{Heath}
T.~ L.~Heath, {\em A History of Greek Mathematics}, Oxford (1921).

\bibitem{Goldstein}
B.~R.~Goldstein, 
Historia Math.~11 (4), 411 (1984).

\bibitem{Rawlins1}
D.~Rawlins, 
Isis 73, 259 (1982).

\bibitem{Rawlins2}
D.~Rawlins, 
Arch. Hist. Exact Sci.~26 (3), 211 (1982).

\bibitem{Gulbekian}
E.~Gulbekian, 
Arch. Hist. Exact Sci.~37 (4), 359 (1987).

\bibitem{reviews}
See, for example,
  G.~Jungman, M.~Kamionkowski and K.~Griest,
  Phys.\ Rept.\  {\bf 267}, 195 (1996)
  [hep-ph/9506380];
  L.~Bergstrom,
  Rept.\ Prog.\ Phys.\  {\bf 63}, 793 (2000)
  [hep-ph/0002126];
  J.~L.~Feng,
  eConf {\bf C0307282}, L11 (2003)
  [hep-ph/0405215];
  Annals Phys.\  {\bf 315}, 2 (2005);
  G.~Bertone, D.~Hooper and J.~Silk,
  Phys.\ Rept.\  {\bf 405}, 279 (2005)
  [hep-ph/0404175];
  M.~Drees, R.~Godbole and P.~Roy,
  {\em Theory and Phenomenology of Sparticles: An Account of
  Four-dimensional $N=1$ Supersymmetry in High Energy Physics}, World
  Scientific (2004).

\bibitem{Peccei:1977ur}
R.~D.~Peccei and H.~R.~Quinn,
Phys.\ Rev.\ D {\bf 16}, 1791 (1977).

\bibitem{Wilczek:pj}
F.~Wilczek,
Phys.\ Rev.\ Lett.\  {\bf 40}, 279 (1978).

\bibitem{Weinberg:1977ma}
S.~Weinberg,
Phys.\ Rev.\ Lett.\  {\bf 40}, 223 (1978).

\bibitem{Pagels:ke}
H.~Pagels and J.~R.~Primack,
Phys.\ Rev.\ Lett.\  {\bf 48}, 223 (1982).

\bibitem{Weinberg:zq}
S.~Weinberg,
Phys.\ Rev.\ Lett.\  {\bf 48}, 1303 (1982).

\bibitem{Krauss:1983ik}
L.~M.~Krauss,
Nucl.\ Phys.\ B {\bf 227}, 556 (1983).

\bibitem{Nanopoulos:1983up}
D.~V.~Nanopoulos, K.~A.~Olive and M.~Srednicki,
Phys.\ Lett.\ B {\bf 127}, 30 (1983).

\bibitem{Khlopov:pf}
M.~Y.~Khlopov and A.~D.~Linde,
Phys.\ Lett.\ B {\bf 138} (1984) 265.

\bibitem{Ellis:1984eq}
J.~R.~Ellis, J.~E.~Kim and D.~V.~Nanopoulos,
Phys.\ Lett.\ B {\bf 145}, 181 (1984).

\bibitem{Ellis:1984er}
J.~R.~Ellis, D.~V.~Nanopoulos and S.~Sarkar,
Nucl.\ Phys.\ B {\bf 259}, 175 (1985).

\bibitem{Juszkiewicz:gg}
R.~Juszkiewicz, J.~Silk and A.~Stebbins,
Phys.\ Lett.\ B {\bf 158}, 463 (1985).

\bibitem{Goldberg:1983nd}
H.~Goldberg,
Phys.\ Rev.\ Lett.\  {\bf 50}, 1419 (1983).

\bibitem{Ellis:1983ew}
  J.~R.~Ellis, J.~S.~Hagelin, D.~V.~Nanopoulos, K.~A.~Olive and
  M.~Srednicki,
  Nucl.\ Phys.\ B {\bf 238}, 453 (1984).

\bibitem{Rajagopal:1990yx}
  K.~Rajagopal, M.~S.~Turner and F.~Wilczek,
  Nucl.\ Phys.\ B {\bf 358}, 447 (1991).

\bibitem{Kusenko:1997si}
A.~Kusenko and M.~E.~Shaposhnikov,
Phys.\ Lett.\ B {\bf 418}, 46 (1998)
[hep-ph/9709492].

\bibitem{Chung:1998ua}
D.~J.~H.~Chung, E.~W.~Kolb and A.~Riotto,
Phys.\ Rev.\ Lett.\  {\bf 81}, 4048 (1998)
[hep-ph/9805473].

\bibitem{Spergel:1999mh}
D.~N.~Spergel and P.~J.~Steinhardt,
Phys.\ Rev.\ Lett.\  {\bf 84}, 3760 (2000)
[astro-ph/9909386].

\bibitem{Kaplinghat:2000vt}
M.~Kaplinghat, L.~Knox and M.~S.~Turner,
Phys.\ Rev.\ Lett.\  {\bf 85}, 3335 (2000)
[astro-ph/0005210].

\bibitem{Servant:2002aq}
G.~Servant and T.~M.~P.~Tait,
Nucl.\ Phys.\ B {\bf 650}, 391 (2003)
[hep-ph/0206071].

\bibitem{Cheng:2002ej}
H.~C.~Cheng, J.~L.~Feng and K.~T.~Matchev,
Phys.\ Rev.\ Lett.\  {\bf 89}, 211301 (2002)
[hep-ph/0207125].

\bibitem{Cembranos:2003mr}
J.~A.~R.~Cembranos, A.~Dobado and A.~L.~Maroto,
Phys.\ Rev.\ Lett.\  {\bf 90}, 241301 (2003)
[hep-ph/0302041].

\bibitem{Cembranos:2003fu}
J.~A.~R.~Cembranos, A.~Dobado and A.~L.~Maroto,
Phys.\ Rev.\ D {\bf 68}, 103505 (2003)
[hep-ph/0307062].

\bibitem{Feng:2003xh}
J.~L.~Feng, A.~Rajaraman and F.~Takayama,
Phys.\ Rev.\ Lett.\  {\bf 91}, 011302 (2003)
[hep-ph/0302215].

\bibitem{Feng:2003uy}
J.~L.~Feng, A.~Rajaraman and F.~Takayama,
Phys.\ Rev.\ D {\bf 68}, 063504 (2003)
[hep-ph/0306024].

\bibitem{discovering}
HEPAP LHC/ILC Subpanel, ``Discovering the Quantum Universe,''
http://www.linearcollider.org.

\bibitem{Cheng:2003ju}
  H.~C.~Cheng and I.~Low,
  JHEP {\bf 0309}, 051 (2003)
  [hep-ph/0308199].

\bibitem{whitepaper} 
Report of the Cosmology Subgroup, American Linear Collider Physics
Group, in preparation.

\bibitem{Gray:2005ci}
  R.~Gray {\it et al.},
  hep-ex/0507008.

\bibitem{Birkedal:2005jq}
  A.~Birkedal {\it et al.},
  hep-ph/0507214.

\bibitem{Battaglia:2005ie}
  M.~Battaglia and M.~E.~Peskin,
  hep-ph/0509135.

\bibitem{Allanach:2004xn}
  B.~C.~Allanach, G.~Belanger, F.~Boudjema and A.~Pukhov,
  JHEP {\bf 0412}, 020 (2004)
  [hep-ph/0410091].

\bibitem{Moroi:2005nc}
  T.~Moroi, Y.~Shimizu and A.~Yotsuyanagi,
  Phys.\ Lett.\ B {\bf 625}, 79 (2005)
  [hep-ph/0505252].

\bibitem{Paige:2003mg}
  F.~E.~Paige, S.~D.~Protopescu, H.~Baer and X.~Tata,
  hep-ph/0312045.

\bibitem{Gondolo:2004sc}
  P.~Gondolo, J.~Edsjo, P.~Ullio, L.~Bergstrom, M.~Schelke and
  E.~A.~Baltz,
  JCAP {\bf 0407}, 008 (2004)
  [astro-ph/0406204].

\bibitem{Belanger:2004yn}
  G.~Belanger, F.~Boudjema, A.~Pukhov and A.~Semenov,
  hep-ph/0405253.

\bibitem{anngraphs} 
These Feynman graphs are presented in the first paper of
Ref.~\protect\cite{reviews}, and are analyzed in full detail in
  M.~Drees and M.~M.~Nojiri,
  Phys.\ Rev.\ D {\bf 47}, 376 (1993)
  [hep-ph/9207234];
  M.~Drees, G.~Jungman, M.~Kamionkowski and M.~M.~Nojiri,
  Phys.\ Rev.\ D {\bf 49}, 636 (1994)
  [hep-ph/9306325].

\bibitem{Allanach:2002nj}
  B.~C.~Allanach {\it et al.},
in {\it Proc. of the APS/DPF/DPB Summer Study on the Future of
  Particle Physics (Snowmass 2001) } ed. N.~Graf,
  Eur.\ Phys.\ J.\ C {\bf 25}, 113 (2002)
  [eConf {\bf C010630}, P125 (2001)]
  [hep-ph/0202233].

\bibitem{Weiglein:2004hn}
  G.~Weiglein {\it et al.}  [LHC/LC Study Group],
  hep-ph/0410364.

\bibitem{Feng:2001ce}
  J.~L.~Feng and M.~E.~Peskin,
  Phys.\ Rev.\ D {\bf 64}, 115002 (2001)
  [hep-ph/0105100].

\bibitem{Feng:1998ud}
  J.~L.~Feng,
  Int.\ J.\ Mod.\ Phys.\ A {\bf 13}, 2319 (1998)
  [hep-ph/9803319];
  Int.\ J.\ Mod.\ Phys.\ A {\bf 15}, 2355 (2000)
  [hep-ph/0002055].

\bibitem{Freitas:2003yp}
  A.~Freitas, A.~von Manteuffel and P.~M.~Zerwas,
  Eur.\ Phys.\ J.\ C {\bf 34}, 487 (2004)
  [hep-ph/0310182].

\bibitem{Feng:1999hg}
  J.~L.~Feng and T.~Moroi,
  Phys.\ Rev.\ D {\bf 61}, 095004 (2000)
  [hep-ph/9907319].

\bibitem{Feng:1999mn}
  J.~L.~Feng, K.~T.~Matchev and T.~Moroi,
  Phys.\ Rev.\ Lett.\  {\bf 84}, 2322 (2000)
  [hep-ph/9908309];
  Phys.\ Rev.\ D {\bf 61}, 075005 (2000)
  [hep-ph/9909334].

\bibitem{Baer:2005ky}
  H.~Baer, T.~Krupovnickas, S.~Profumo and P.~Ullio,
  hep-ph/0507282.

\bibitem{Feng:2000gh}
  J.~L.~Feng, K.~T.~Matchev and F.~Wilczek,
  Phys.\ Lett.\ B {\bf 482}, 388 (2000)
  [hep-ph/0004043];
  Phys.\ Rev.\ D {\bf 63}, 045024 (2001)
  [astro-ph/0008115].

\bibitem{Baer:2003jb}
  H.~Baer, C.~Balazs, A.~Belyaev and J.~O'Farrill,
  JCAP {\bf 0309}, 007 (2003)
  [hep-ph/0305191].

\bibitem{Ellis:2003dn}
  J.~R.~Ellis, K.~A.~Olive, Y.~Santoso and V.~C.~Spanos,
  Phys.\ Lett.\ B {\bf 588}, 7 (2004)
  [hep-ph/0312262].

\bibitem{Feng:2004zu}
  J.~L.~Feng, S.~Su and F.~Takayama,
  Phys.\ Rev.\ D {\bf 70}, 063514 (2004)
  [hep-ph/0404198].

\bibitem{Feng:2004mt}
  J.~L.~Feng, S.~Su and F.~Takayama,
  Phys.\ Rev.\ D {\bf 70}, 075019 (2004)
  [hep-ph/0404231].

\bibitem{Wang:2004ib}
  F.~Wang and J.~M.~Yang,
  Eur.\ Phys.\ J.\ C {\bf 38}, 129 (2004)
  [hep-ph/0405186].

\bibitem{Ellis:2004bx}
  J.~R.~Ellis, K.~A.~Olive, Y.~Santoso and V.~C.~Spanos,
  Phys.\ Lett.\ B {\bf 603}, 51 (2004)
  [hep-ph/0408118].

\bibitem{Roszkowski:2004jd}
  L.~Roszkowski and R.~Ruiz de Austri,
  JHEP {\bf 0508}, 080 (2005)
  [hep-ph/0408227].

\bibitem{axinos}
  L.~Covi, J.~E.~Kim and L.~Roszkowski,
  Phys.\ Rev.\ Lett.\  {\bf 82}, 4180 (1999)
  [hep-ph/9905212];
   L.~Covi, H.~B.~Kim, J.~E.~Kim and L.~Roszkowski,
   JHEP {\bf 0105}, 033 (2001)
   [hep-ph/0101009];
  L.~Covi, L.~Roszkowski, R.~Ruiz de Austri and M.~Small,
  JHEP {\bf 0406}, 003 (2004)
  [hep-ph/0402240].

\bibitem{Bi:2003qa}
  X.~J.~Bi, M.~z.~Li and X.~m.~Zhang,
  Phys.\ Rev.\ D {\bf 69}, 123521 (2004)
  [hep-ph/0308218].

\bibitem{Feng:2003nr}
J.~L.~Feng, A.~Rajaraman and F.~Takayama,
Phys.\ Rev.\ D {\bf 68}, 085018 (2003)
[hep-ph/0307375].

\bibitem{Kitano:2005ge}
  R.~Kitano and I.~Low,
  hep-ph/0503112.

\bibitem{Ellis:1990nb}
J.~R.~Ellis, G.~B.~Gelmini, J.~L.~Lopez, D.~V.~Nanopoulos and S.~Sarkar,
Nucl.\ Phys.\ B {\bf 373}, 399 (1992).

\bibitem{Moroi:1993mb} 
T.~Moroi, H.~Murayama and M.~Yamaguchi,
Phys.\ Lett.\ B {\bf 303}, 289 (1993).

\bibitem{Bolz:2000fu}
M.~Bolz, A.~Brandenburg and W.~Buchmuller,
Nucl.\ Phys.\ B {\bf 606}, 518 (2001)
[hep-ph/0012052].

\bibitem{Brandenburg:2004du}
Other production mechanisms for axinos have also been considered in
  A.~Brandenburg and F.~D.~Steffen,
  JCAP {\bf 0408}, 008 (2004)
  [hep-ph/0405158].

\bibitem{Burles:2000zk}
S.~Burles, K.~M.~Nollett and M.~S.~Turner,
Astrophys.\ J.\  {\bf 552}, L1 (2001)
[astro-ph/0010171].

\bibitem{Thorburn}
J.~A.~Thorburn,
Astrophys.\ J.\  {\bf 421}, 318 (1994).

\bibitem{Bonafacio}
P.~Bonifacio and P.~Molaro,
MNRAS, {\bf 285}, 847 (1997).

\bibitem{Ryan:1999vr}
S.~G.~Ryan, T.~C.~Beers, K.~A.~Olive, B.~D.~Fields and J.~E.~Norris,
Astrophys.\ J.\ Lett. {\bf 530}, L57 (2000)
[astro-ph/9905211].

\bibitem{Pinsonneault:1998nf}
M.~H.~Pinsonneault, T.~P.~Walker, G.~Steigman and V.~K.~Narayanan,
Astrophys.\ J. {\bf 527}, 180 (1999)
[astro-ph/9803073].

\bibitem{Vauclair:1998it}
S.~Vauclair and C.~Charbonnel,
Astrophys.\ J. {\bf 502}, 372 (1998)
[astro-ph/9802315].

\bibitem{Kawasaki:1994sc}
M.~Kawasaki and T.~Moroi,
Astrophys.\ J.\  {\bf 452}, 506 (1995)
[astro-ph/9412055].

\bibitem{Holtmann:1998gd}
E.~Holtmann, M.~Kawasaki, K.~Kohri and T.~Moroi,
Phys.\ Rev.\ D {\bf 60}, 023506 (1999)
[hep-ph/9805405].

\bibitem{Kawasaki:2000qr}
M.~Kawasaki, K.~Kohri and T.~Moroi,
Phys.\ Rev.\ D {\bf 63}, 103502 (2001)
[hep-ph/0012279].

\bibitem{Cyburt:2002uv}
  R.~H.~Cyburt, J.~R.~Ellis, B.~D.~Fields and K.~A.~Olive,
  Phys.\ Rev.\ D {\bf 67}, 103521 (2003)
  [astro-ph/0211258].

\bibitem{Ellis:2005ii}
  J.~R.~Ellis, K.~A.~Olive and E.~Vangioni,
  Phys.\ Lett.\ B {\bf 619}, 30 (2005)
  [astro-ph/0503023].

\bibitem{Cerdeno:2005eu}
See, for example,
  D.~G.~Cerdeno, K.~Y.~Choi, K.~Jedamzik, L.~Roszkowski and R.~Ruiz de Austri,
  hep-ph/0509275.

\bibitem{Reno:1987qw}
M.~H.~Reno and D.~Seckel,
Phys.\ Rev.\ D {\bf 37}, 3441 (1988).

\bibitem{Dimopoulos:1987fz}
S.~Dimopoulos, R.~Esmailzadeh, L.~J.~Hall and G.~D.~Starkman,
Astrophys.\ J.\  {\bf 330}, 545 (1988).

\bibitem{Dimopoulos:1988ue}
S.~Dimopoulos, R.~Esmailzadeh, L.~J.~Hall and G.~D.~Starkman,
Nucl.\ Phys.\ B {\bf 311}, 699 (1989).

\bibitem{Kohri:2001jx}
K.~Kohri,
Phys.\ Rev.\ D {\bf 64}, 043515 (2001)
[astro-ph/0103411].

\bibitem{Jedamzik:2004er}
  K.~Jedamzik,
  Phys.\ Rev.\ D {\bf 70}, 063524 (2004)
  [astro-ph/0402344].

\bibitem{Kawasaki:2004yh}
  M.~Kawasaki, K.~Kohri and T.~Moroi,
  Phys.\ Lett.\ B {\bf 625}, 7 (2005)
  [astro-ph/0402490];
  Phys.\ Rev.\ D {\bf 71}, 083502 (2005)
  [astro-ph/0408426].

\bibitem{Hu:1993gc}
  W.~Hu and J.~Silk,
  Phys.\ Rev.\ Lett.\  {\bf 70}, 2661 (1993).

\bibitem{Lamon:2005jc}
  R.~Lamon and R.~Durrer,
  hep-ph/0506229.

\bibitem{Fixsen:1996nj}
  D.~J.~Fixsen, E.~S.~Cheng, J.~M.~Gales, J.~C.~Mather, R.~A.~Shafer
  and E.~L.~Wright,
  Astrophys.\ J.\  {\bf 473}, 576 (1996)
  [astro-ph/9605054].

\bibitem{Eidelman:2004wy}
S.~Eidelman {\it et al.}  [Particle Data Group Collaboration],
Phys.\ Lett.\ B {\bf 592}, 1 (2004).

\bibitem{ARCADE}
http://arcade.gsfc.nasa.gov/arcade;
http://map.gsfc.nasa.gov/DIMES.

\bibitem{Drees:1990yw}
M.~Drees and X.~Tata,
Phys.\ Lett.\ B {\bf 252}, 695 (1990).

\bibitem{Goity:1993ih}
J.~L.~Goity, W.~J.~Kossler and M.~Sher,
Phys.\ Rev.\ D {\bf 48}, 5437 (1993)
[hep-ph/9305244].

\bibitem{Nisati:1997gb}
A.~Nisati, S.~Petrarca and G.~Salvini,
Mod.\ Phys.\ Lett.\ A {\bf 12}, 2213 (1997)
[hep-ph/9707376].

\bibitem{Feng:1997zr} 
J.~L.~Feng and T.~Moroi,
Phys.\ Rev.\ D {\bf 58}, 035001 (1998) 
[hep-ph/9712499].

\bibitem{Hamaguchi:2004df}
  K.~Hamaguchi, Y.~Kuno, T.~Nakaya and M.~M.~Nojiri,
  Phys.\ Rev.\ D {\bf 70}, 115007 (2004)
  [hep-ph/0409248].

\bibitem{Feng:2004yi}
  J.~L.~Feng and B.~T.~Smith,
  Phys.\ Rev.\ D {\bf 71}, 015004 (2005)
  [hep-ph/0409278].

\bibitem{Brandenburg:2005he}
  A.~Brandenburg, L.~Covi, K.~Hamaguchi, L.~Roszkowski and F.~D.~Steffen,
  Phys.\ Lett.\ B {\bf 617}, 99 (2005)
  [hep-ph/0501287].

\bibitem{DeRoeck:2005bw}
  A.~De Roeck, J.~R.~Ellis, F.~Gianotti, F.~Moortgat, K.~A.~Olive and
  L.~Pape,
  hep-ph/0508198.

\bibitem{Buchmuller:2004rq}
W.~Buchmuller, K.~Hamaguchi, M.~Ratz and T.~Yanagida,
Phys.\ Lett.\ B {\bf 588}, 90 (2004)
[hep-ph/0402179].

\bibitem{Feng:2004gn}
  J.~L.~Feng, A.~Rajaraman and F.~Takayama,
  Int.\ J.\ Mod.\ Phys.\ D {\bf 13}, 2355 (2004)
  [hep-th/0405248].

\bibitem{Moore:1994yx}
  B.~Moore,
  Nature {\bf 370}, 629 (1994);
  R.~A.~Flores and J.~R.~Primack,
  Astrophys.\ J.\  {\bf 427}, L1 (1994)
  [astro-ph/9402004];
  J.~J.~Binney and N.~W.~Evans,
  Mon.\ Not.\ Roy.\ Astron.\ Soc.\  {\bf 327}, L27 (2001)
  [astro-ph/0108505];
   A.~R.~Zentner and J.~S.~Bullock,
   Phys.\ Rev.\ D {\bf 66}, 043003 (2002)
   [astro-ph/0205216];
  J.~D.~Simon  {\it et al.},
  Astrophys.\ J.\  {\bf 621}, 757 (2005)
  [astro-ph/0412035].

\bibitem{Klypin:1999uc}
  A.~A.~Klypin, A.~V.~Kravtsov, O.~Valenzuela and F.~Prada,
  Astrophys.\ J.\  {\bf 522}, 82 (1999)
  [astro-ph/9901240];
   A.~R.~Zentner and J.~S.~Bullock,
   Astrophys.\ J.\  {\bf 598}, 49 (2003)
   [astro-ph/0304292].

\bibitem{Sigurdson:2003vy}
  K.~Sigurdson and M.~Kamionkowski,
  Phys.\ Rev.\ Lett.\  {\bf 92}, 171302 (2004)
  [astro-ph/0311486].

\bibitem{Profumo:2004qt}
  S.~Profumo, K.~Sigurdson, P.~Ullio and M.~Kamionkowski,
  Phys.\ Rev.\ D {\bf 71}, 023518 (2005)
  [astro-ph/0410714].

\bibitem{Kaplinghat:2005sy}
  M.~Kaplinghat,
  Phys.\ Rev.\ D {\bf 72}, 063510 (2005)
  [astro-ph/0507300].

\bibitem{Cembranos:2005us}
  J.~A.~R.~Cembranos, J.~L.~Feng, A.~Rajaraman and F.~Takayama,
  Phys.\ Rev.\ Lett.\  {\bf 95}, 181301 (2005)
  [hep-ph/0507150].

\bibitem{Jedamzik:2005sx}
  K.~Jedamzik, M.~Lemoine and G.~Moultaka,
  astro-ph/0508141.

\end{thebibliography}
\end{document}